# Quantitative Evaluation of Common Cause Failures in High Safety-significant Safety-related Digital Instrumentation and Control Systems in Nuclear Power Plants

Han Bao [a]*, Hongbin Zhang [b], Tate Shorthill [c], Edward Chen [d], Svetlana Lawrence [e]

[a] Idaho National Laboratory, P.O. Box 1625, MS 3860, Idaho Falls, ID 83415
[b] Terrapower, 15800 Northup Way, Bellevue, WA 98008
[c] Department of Mechanical Engineering and Materials Science, University of Pittsburgh, 3700 O'Hara Street, Pittsburgh, PA 15261
[d] Department of Nuclear Engineering, North Carolina State University, 2500 Stinson Dr., Raleigh, NC 27607
[e] Idaho National Laboratory, P.O. Box 1625, MS 3860, Idaho Falls, ID 83415

## Abstract

Digital instrumentation and control (DI&C) systems at nuclear power plants (NPPs) have many advantages over analog systems. They are proven to be more reliable, cheaper, and easier to maintain given obsolescence of analog components. However, they also pose new engineering and technical challenges, such as possibility of common cause failures (CCFs) unique to digital systems. This paper proposes a Platform for Risk Assessment of DI&C (PRADIC) that is developed by Idaho National Laboratory (INL). A methodology for evaluation of software CCFs in high safety-significant safety-related DI&C systems of NPPs was developed as part of the framework. The framework integrates three stages of a typical risk assessment—qualitative hazard analysis and quantitative reliability and consequence analyses. The quantified risks compared with respective acceptance criteria provide valuable insights for system architecture alternatives allowing design optimization in terms of risk reduction and cost savings. A comprehensive case study performed to demonstrate the framework's capabilities is documented in this paper. Results show that the PRADIC is a powerful tool capable to identify potential digital-based CCFs, estimate their probabilities, and evaluate their impacts on system and plant safety.

## 1. Introduction

Although the current fleet of the U.S. nuclear power plants (NPPs) was originally designed and constructed with analog systems, the U.S. nuclear industry has been working on transitioning from analog to digital instrumentation and control (DI&C) technologies. DI&C systems have many advantages over analog systems. They are proven to be more reliable, cheaper, and easier to maintain given obsolescence of analog components. However, they also pose new engineering and technical challenges. The U.S. Nuclear Regulatory Commission (NRC) continues supporting the research work in developing and improving licensing criteria for the evaluation of new DI&C systems. In 2018, SECY-18-0090 [1] was published to clarify guidance associated with evaluating potential common cause failures (CCFs) of DI&C systems. The SECY-18-0090 identifies these guiding principles: applicants and licensees for production and utilization facilities should continue to assess and address CCFs due to software for DI&C systems and components; a defense-in-depth and diversity (D3) analysis for reactor trip systems (RTSs), and engineered safety features should continue to be performed to demonstrate that vulnerabilities to a CCF have been identified and adequately addressed. The D3 analysis can be performed using either a design-basis deterministic approach or best-estimate approach [1]. In 2019, the NRC staff developed the integrated action plan (IAP) [2]. Four detailed modernization plans were proposed to resolve regulatory challenges, provide confidence to licensees, and modernize the I&C regulatory infrastructure. One of them—protection against CCF—addresses "developing guidance for using effective qualitative assessments of the likelihood of failures, along with coping and/or bounding analysis for addressing CCFs, use of defensive design measures for eliminating CCF from further consideration, and staff evaluation of the NRC's existing positions on defense against CCF." The current guidance, however, is unclear regarding the applicability of criteria for using coping analysis and other design features (e.g., defensive measures) for eliminating CCFs from further consideration [2]. Meanwhile, the industry stakeholders are seeking clearer NRC staff guidance on methods for analysis of the potential for CCFs in DI&C systems and a more risk-informed, consequence-based regulatory infrastructure that removes uncertainty in requirements and enables technical consistency [2].

Many efforts from regulatory, industrial, and academic communities have been made for qualitatively addressing CCFs in DI&C Systems, especially software CCFs, given the increased pace of design and deployment of high safety-significant safety-related (HSSSR) DI&C systems in NPPs. To successfully model DI&C systems, the need exists to model both the hardware and software interactions of the system. Traditional methods, such as failure modes and effects analysis (FMEA) and fault tree analysis (FTA), have been used to extensively model analog systems. However, these traditional methods are not fully suitable to identify failures in interactions between digital systems and controlled processes (i.e., Type 1 interactions) as well as interactions between digital systems and their own components or other systems (i.e., Type 2 interactions) [3]. Lessons learned from the NRC's investigation of multiple analysis methods indicate there "may not be one preferable method for modeling all digital systems" [3]. Combining methods may prove beneficial. A recent advancement in hazards analysis, developed jointly by Electric Power Research Institute (EPRI) and Sandia National Laboratory, combines FTA and the systems-theoretic process analysis (STPA) as a portion of their methodology for Hazard and Consequence Analysis for Digital Systems (HAZCADS) [4]. Though STPA may be applied at any stage of system design and review, it is ideally suited for early applications in the design process before safety features have been incorporated into the design [5]. Then, as more details are incorporated, the STPA method is applied iteratively to further improve the design. However, even when fine details about a system are known, the analysis may remain at a high level, relying on causal factor investigations to provide details of subcomponent failures and interactions. In other words, details, such as redundant subsystems or components, are often ignored in all but the final part of STPA.

In July 2021, Nuclear Energy Institute (NEI) published NEI 20-07, "Guidance for Addressing Software Common Cause Failure in High Safety-Significant Safety-Related Digital I&C Systems" [6]. A two-step process was proposed to address HSSSR systematic CCFs based on STPA: Step 1 is to perform a systematic hazards analysis based on STPA that creates a model of the system control structure, identifies unsafe control actions (UCAs) as software failures, and establishes a risk reduction objective (RRO); Step 2 is to develop STPA loss scenarios and eliminate or mitigate them in an efficient way. A bounding assessment is proposed to calculate the risk change when entire HSSSR systems fail due to software CCFs (assuming system failure probability = 1). The risk change (e.g., $\Delta$ core damage frequency [CDF]) is then mapped to the regions described in RG 1.174 [7] and used to determine the RRO. This process qualitatively addresses potential failures in DI&C based on a bounding assessment; consequently, the real safety margin gained by plant digitalization on HSSSR DI&C systems could be underestimated in this intentionally conservative approach.

The described above efforts provides a technical basis for dealing with potential software CCFs in the HSSSR DI&C systems of NPPs; however, some technical challenges remain:
1. Is qualitative evaluation sufficient for addressing software CCFs in HSSSR DI&C systems? Most of the STPA-based approaches mentioned above focus on the identification of software failures but not the quantification of their probabilities. Although these software failures are added into an integrated fault tree (FT), their probabilities are not calculated. Instead, a conservative bounding assessment is performed to evaluate their impacts to plant safety (e.g., $\Delta$ CDF), which may lead to an underestimation of safety margins gained by plant digitalization and/or skewed risk metrics.

2. How to quantitatively evaluate CCF-related impacts to HSSSR DI&C systems and entire plant response? This proposes a need in developing an integrated strategy to include both qualitative hazard analysis and quantitative reliability and consequence analysis for addressing software CCF issues in HSSSR DI&C systems. Inputs and outputs of each analysis process should be consistently connected.
3. How to efficiently identify the most significant CCFs, especially software CCFs? Existing STPA-based approaches represents good performance in capturing systematic failures in digital interactions; however, there is no clear representation of how to create a control structure for a complicated system containing multiple layers of redundancy and diversity.
4. How to perform a complete reliability analysis for large-scale HSSSR DI&C systems with small-scale software/digital units? Currently, there is no consensus method for the software reliability modeling of digital systems in NPPs. A reliability analysis approach is needed, especially for the quantification of UCAs from STPA analyses.
5. How to evaluate different system architectures from perspectives of both risk and cost? A DI&C system could be designed using several options for system architecture (e.g., redundancy and diversity at different system levels), and a comprehensive, consistent, integrated approach is needed to support evaluation of various design architectures to ensure the most optimal one is selected for implementation. This approach should be able to support vendors and utilities with optimization of design solutions from economical perspectives given the constrain of meeting risk-informed safety requirements.
6. Lastly, a need clearly exists to develop a risk assessment strategy to support quantitative D3 analyses for assuring the long-term safety and reliability of vital digital systems and reducing uncertainties in costs, time, and supporting integration of digital systems during the design stage of the plant.

To address the above-mentioned challenges, an integrated risk assessment strategy is needed to include both qualitative hazard analysis and quantitative reliability and consequence analysis for addressing software CCF in the HSSSR DI&C systems. To fulfill this need and deal with the technical barriers in identifying potential software CCF issues in HSSSR DI&C systems of NPPs, the Risk-Informed Systems Analysis (RISA) Pathway sponsored by the U.S. Department of Energy (DOE) Light Water Reactor Sustainability (LWRS) Program, initiated a project to develop a risk assessment strategy [8] to:

1. Provide a best-estimate, risk-informed capability to quantitatively and accurately estimate the NPP safety margin gained from modernization of HSSSR DI&C systems
2. Develop an advanced risk assessment technology to support the transition from analog to DI&C technologies for nuclear industry
3. Assure the long-term safety and reliability of HSSSR DI&C systems
4. Reduce uncertainty in costs and support integration of DI&C systems at NPPs.

A Platform for Risk Assessment of DI&C (PRADIC, previously defined as IRADIC) was proposed to address the project objectives. As shown in Figure 1, the proposed framework provides a systematic, verifiable, and reproducible approach based on technically sound methodologies. The framework successively implements qualitative hazard analysis, quantitative reliability analysis, and consequence analysis to obtain quantitative risk metrics. The quantified risks are then compared with respective acceptance criteria which allows identification of vulnerabilities as well as to provide suggestions for risk reduction and design optimization.

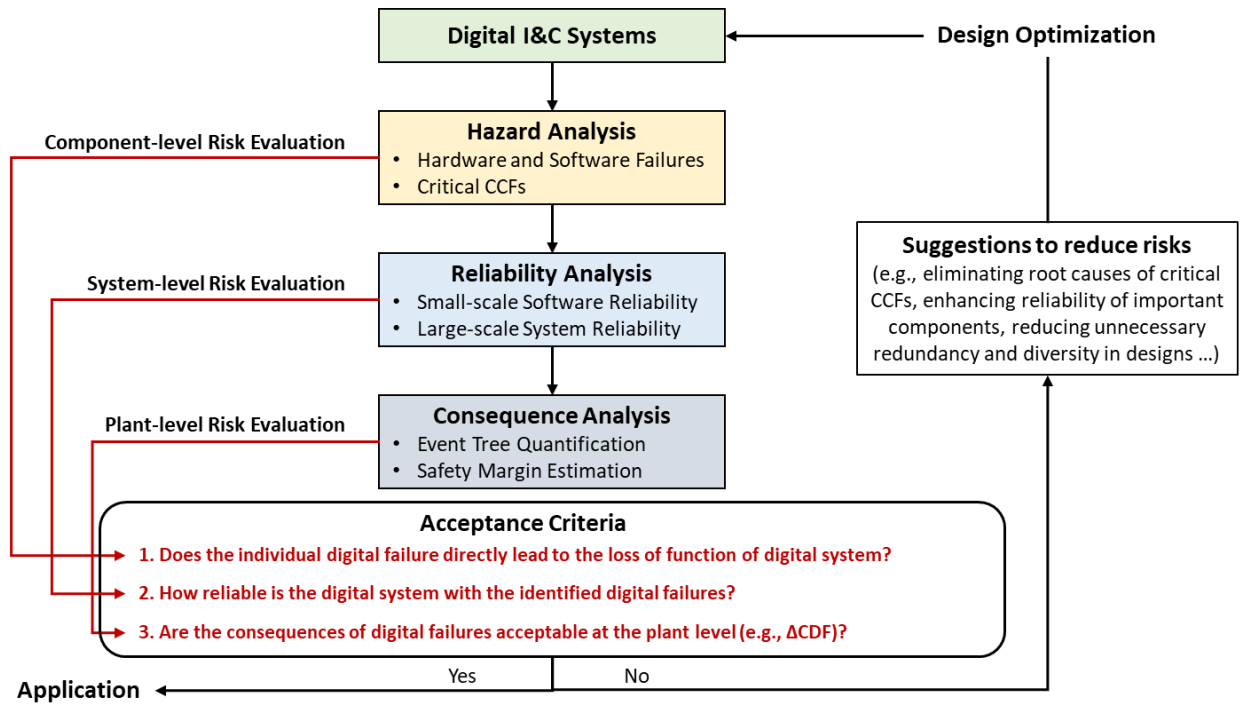

Figure 1. Schematic of the PRADIC for safety evaluation and design optimization of HSSSR DI&C systems (derived from [9]).

Section 2 introduces the value proposition of the INL-PRADIC. Sections 3 and 4, respectively, describe the technical approach via case studies performed for representative HSSSR DI&C systems. Section 5 documents the consequence analysis performed for a generic pressurized-water reactor (PWR). Section 6 provides conclusions and outlines future work.

## 2. Value Proposition of INL-PRADIC

To deal with the technical issues in addressing potential software CCF issues in HSSSR DI&C systems of NPPs, the PRADIC is expected to provide:

1. An integrated and best-estimate, risk-informed capability to address new technical digital issues quantitatively, accurately, and efficiently in plan modernization progress, such as software CCFs in HSSSR DI&C systems of NPPs.

Existing qualitative approach for addressing CCFs in HSSSR DI&C systems may significantly underestimate the real safety margin introduced by plant digitalization. The PRADIC is developed and demonstrated in an integrated way including both qualitative hazard analysis and quantitative reliability and consequence analyses. The PRADIC aims to provide a best-estimate, risk-informed capability to accurately estimate the safety margin increase obtained from plant modernization, especially for the digital HSSSR I&C systems.

In the PRADIC, a redundancy-guided systems-theoretic method for hazard analysis (RESHA) was developed HSSSR DI&C systems for supporting I&C designers and engineers to address both hardware and software CCFs and qualitatively analyze their effects on system availability [10] [11]. It also provides a technical basis for implementing, following reliability and consequence analyses of unexpected software failures, and supporting the optimization of D3 applications in a cost-efficient way. Targeting the complexity of redundant designs in HSSSR DI&C systems integrates STPA [5], FTA, and HAZCADS [4] to effectively identify software CCFs by reframing STPA in a redundancy-guided way, such as (1) depicting a redundant and diverse system via a detailed representation; (2) refining different redundancy levels based on the structure of DI&C systems; (3) constructing a redundancy-guided multilayer control structure; and (4) identifying potential CCFs in different redundancy levels. This approach has been demonstrated and applied for the hazard analysis of a four-division digital RTS [10] and a four-division, digital, engineered safety features actuation system (ESFAS) [11]. These efforts have been included in the LWRS-RISA milestone report for FY 2020 [12] [13].

The second part in risk analysis is reliability analysis with the tasks of (1) quantifying the probabilities of basic events of the integrated FT from the hazard analysis; (2) estimating the probabilities of the consequences of digital system failures. In the PRADIC, two methods have been developed for different application conditions: the Bayesian and Human-reliability-analysis-aided Method for the reliability Analysis of Software (BAHAMAS) [14] for limited-data conditions and Orthogonal-defect Classification for Assessing Software reliability (ORCAS) for data-rich analysis. More details can be found in Section 4.

Finally, consequence analysis is conducted to quantitatively evaluate the consequence impact of digital failures on plant behaviors and responses. Some digital-based failures may initiate an event or scenario that may not be analyzed before, which brings in a big challenge to plant safety. In this paper, a couple of accident scenarios have been selected for the consequence analysis, as described in Section 5.
In February 2022, the NRC organized a public meeting inform the industry and solicit external stakeholders' feedback on the NRC's plan to potentially expand the current policy for addressing CCFs for DI&C systems to allow consideration of risk-informed alternatives. The LWRS Program's RISA team presented on providing capabilities to address and fulfill the risk-informed alternatives for evaluation of CCFs in DI&C systems. The NRC staff found the framework interesting from the regulatory point of view since it may be useful to evaluate the impacts of various DI&C design architectures to the overall plant safety.

2. A common and a modularized platform to digital I&C designers, software developers, cybersecurity analysts, and plant engineers to efficiently predict and prevent risk in the early design stage of digital I&C systems.

Many programs/projects were and are being created with various methods/approaches/frameworks generated either for single software reliability analysis or for quantifying the system-level interactions between digital systems or between digital systems and other systems. However, these efforts are rarely targeting software CCFs in HSSSR DI&C systems.

As shown in Figure 2, the PRADIC, as a modularized platform, aims to have a good communication with various small-scale unit-level software reliability analysis methods (e.g., quantitative software reliability methods) and large-scale system-level reliability analysis frameworks (e.g., probabilistic risk assessment [PRA]). RESHA, as a top-down approach, can identify the digital or software failures in the unit-level interactions inside of a digital system, then BAHAMAS and ORCAS can be used to quantify the probability of the STPA-identified software failures based on suitable existing quantitative software reliability methods such as Bayesian networks, test-based, or metric-based methods.

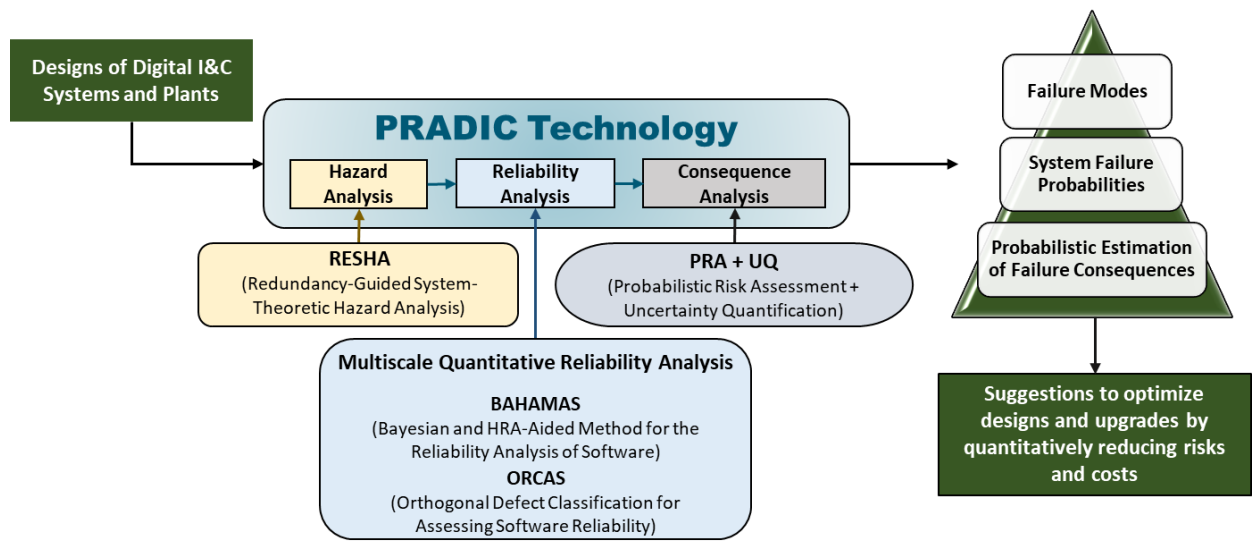

Figure 2. The flexible and modularized structure of the PRADIC.

3. Technical bases and risk-informed insights to assist NRC and industry in formalizing relevant licensing processes for addressing CCF considerations in HSSSR DI&C systems.

Figure 3 illustrates how the PRADIC can support licensing of a HSSSR DI&C design or upgrade. NRC BTP 7-19, "Guidance for Evaluation of Diversity and Defense-In-Depth in Digital Computer-Based Instrumentation and Control Systems Review Responsibilities," [15] clarifies the requirement for acceptable methods for addressing CCFs, including identifying CCFs, reducing CCF likelihood, and evaluating CCF impacts in design-basis events. The capabilities of the PRADIC in hazard, reliability, and consequence analysis matches well with these requirements.

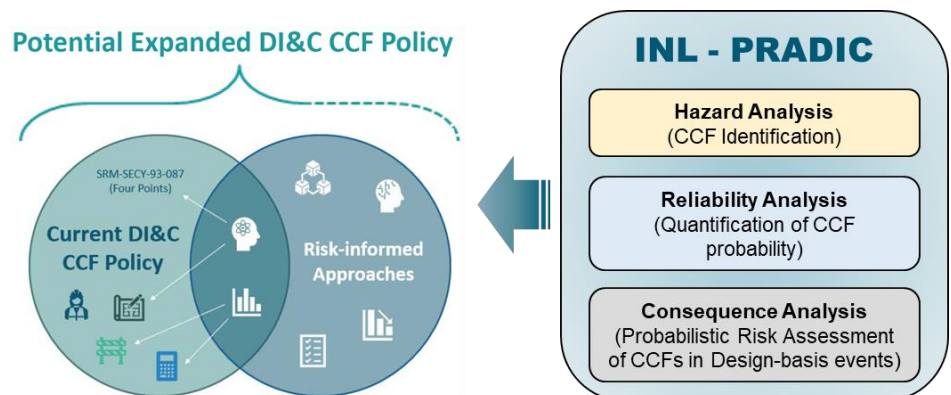

Figure 3. NRC potential expanded DI&C CCF policy vs. INL-PRADIC capability in CCF analysis.

4. An integrated risk-informed tool to support the nuclear industry in addressing regulatory requirements in DI&C system implementation.

The PRADIC can be beneficial for the design of digital HSSSR systems in plant modernization process: the estimated safety margin using the PRADIC should be much higher and more accurate than other conservative bounding assessment approach. The safety improvements of these new digital designs are expected to be significant and can be presented more clearly.

Currently, it is thought after qualitatively addressing CCFs, all of them need to be fixed by adding diversity, which costs a lot. In fact, some of CCFs do not have significant impacts on the change of CDF or large release frequency. The PRADIC can evaluate the impacts of single software CCFs to the HSSSR DI&C systems and even the plant safety, based on which suggestions can be provided to optimize the D3 application in the early design stage of HSSSR DI&C systems. For instance, it can support the determination of the level of redundancy (e.g., a four-division ESFAS vs. a two-division ESFAS) or the level of diversity (e.g., deployment of software/design/equipment diversity in division level vs. in unit level). By comparing the risk and cost of different redundant and diverse designs, cost can be saved if some CCFs are proved to be insignificant to plant safety. Based on current PRADIC analysis results, failure probability of HSSSR DI&C system due to software CCFs is quite low, and the CDF is also significantly reduced compared with the one with traditional analog systems. The PRADIC was also suggested to deal with the software risk analysis for machine-learning-based digital twins in the nearly autonomous management and control systems [16].

## 3. Redundancy-Guided System-Theoretic Hazard Analysis

The hazard analysis method developed for the PRADIC is called "redundancy-guided systems-theoretic method for hazard analysis" (RESHA). The RESHA method is a FTA-based method that incorporates STPA to identify inner software failures and digital-based failures in Type II interactions. Figure 4 illustrates the concepts of Type I and Type II interactions: the interactions of a DI&C system (and/or its components) with a controlled process (e.g., NPPs) and Type II: the interactions of a DI&C system (and/or its components) with itself and/or other digital systems and components [17]. Software should not be analyzed in isolation from the complete digital system. In addition to the inner failures of software, failures in Type II interactions should also be considered in the risk analysis of DI&C system.

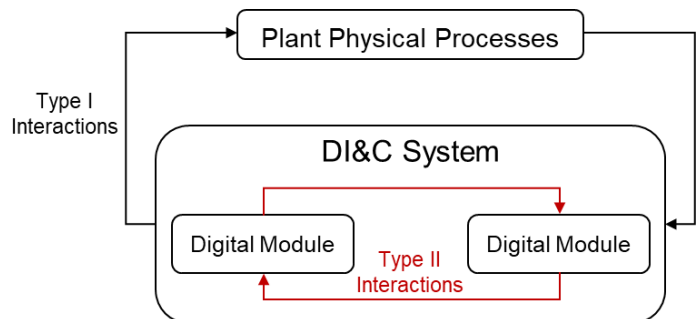

Figure 4. Illustration of Type I between a DI&C system and plant physical processes and Type II interactions between a DI&C system (and/or its components) with itself and/or other digital systems and components.

According to the difference in functionality, there are normally two types of digital modules in DI&C systems: digital controller and intermediate digital module. The physical elements of a digital controller include a central processing unit along with its associated microcode, memory, and I&C program [17]. A digital controller may be connected to other controllers or intermediate modules (e.g., sensors, actuators, or even input/output modules). STPA handbook [5] provides another definition of a controller in an I&C system; a controller provides control actions on the system and gets feedback to determine the impact of the control actions, as shown in Figure 5. The HSSSR DI&C systems in NPPs normally include multiple controllers to ensure the availability and reliability of the actuation of safety controls and features. For example, the rod cluster control assembly can be dropped by manual control from operators or automated control of RTS. In this case, both operators and RTS can be considered as a controller based on the controller definition from STPA. Digital controllers can be defined in different scales for different analysis target. For example, a digital processor inside of a RTS can be treated as a controller when details about Type II interactions inside of RTS are needed, while RTS itself can be a digital controller when details about Type I interactions are needed.

In STPA, process models represent the controller's internal beliefs that are used by the control algorithm to determine control actions. Control algorithms specify how control actions are determined based on the controller's process model, inputs, and feedbacks from previous control processes and intermediate modules. In [14], a process model is thought as the diagnosis portion of a controller, whereas the control algorithm provides actions based on the model's diagnosis.

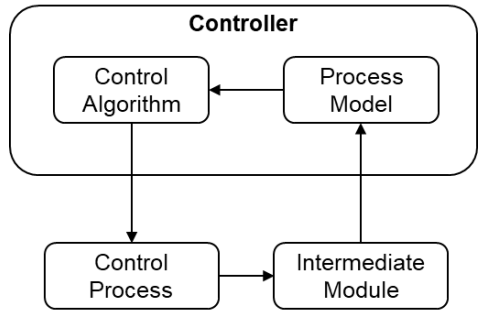

Figure 5. A generic control loop with a controller (derived from STPA handbook [5]).

RESHA identifies UCAs as the digital failures resulting from two categories of causal factors:
1. Category 1: Inner software failure
    a. Software design defects (mainly due to inconsistent process model)
    b. Software implementation defects (mainly due to inadequate control algorithm)
2. Category 2: Misleading, inadequate, or incorrect input conditions
    a. Failures in Type II interactions.

Regarding Category 2 causal factors, RESHA recognizes these as unsafe information flows (UIFs) from dependent intermediate modules. These are explicit indications that there exist digital-based failures from in appropriate Type II interactions. For clarification, an unsafe information flow is regarded as any received signal external to the controller that is erroneous, falsified, or incorrect. In contrast to a UCA, UIFs are failures in the feedback mechanism of the control structure, whereas the former is a failure in the control mechanism. They are considered one of the classes of casual factors for UCAs under the STPA methodology. A causality relationship diagram is provided below to show how UIFs are related to UCAs and associated hazards/losses. Some of the more relevant Type II interactions are listed below but is not exhaustive:

1. Data transmitted by intermediate module is correct but received/interpreted incorrectly at controller
    a. Excessive noise distortion
    b. Data transmission pathway degraded (e.g., loose connecting wire)
    c. Data feedback to controller has incorrect timing/order
2. Data transmitted by intermediate module is not correct
    a. Internal software failure related to design or implementation defects
    b. Data received by the intermediate module from other intermediate modules is incorrect resulting in the output also incorrect.

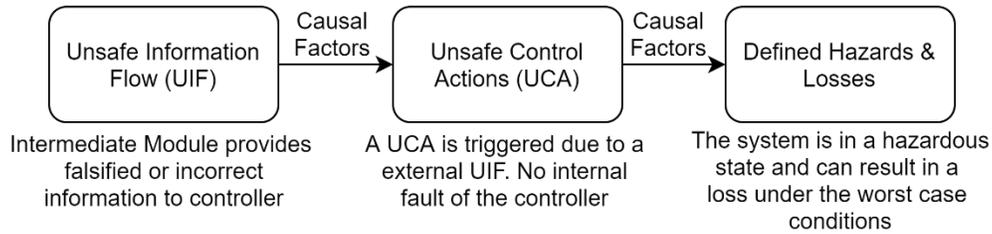

Figure 6. Causality relationship diagram between UIF, UCA, and defined hazards and losses to stakeholders.

### 3.1 Technical Approach

RESHA incorporates the concept of combining FTA and STPA from HAZCADS. STPA is reframed in a redundancy-guided way to address CCF concerns in highly redundant DI&C systems. A seven-step process, as shown in Figure 7, illustrates the workflow of RESHA in the PRADIC for the hazard analysis of DI&C systems, especially for CCF analysis of highly redundant HSSSR DI&C systems. The main outcomes of RESHA are (1) the identification of CCFs and potential single points of failure (SPOFs) in the DI&C design; (2) an integrated FT including both hardware and software failures, individual failures, and CCFs; (3) hazard preventive strategies. The acceptance criterion of risk evaluation for the PRADIC hazard analysis is "[d]oes the individual digital-based failure lead to the loss of function of the DI&C system?" In another words, is there any SPOF existing in the system that may lead to the failure of DI&C system?

RESHA relies on UCAs and an integrated FT to achieve its primary outputs. There are different categories of UCAs (e.g., control action not provided, or control actions provided spuriously) which are integrated with the FT to account for software failures. Depending on the goals of the risk assessment, the integrated FT may also be combined with an event tree (ET) as part of a larger comprehensive risk assessment. For such assessments, the ET links with FT top events which, in turn, guide the selection of UCA types to be used within the FT itself [9]. For example, a top event pertaining to failure on demand will contain UCAs that match that same category. More details about the RESHA methodology can be found in [10] and [11]. Ultimately, UCAs play a large role for hazard analysis within RESHA and PRADIC as a whole.

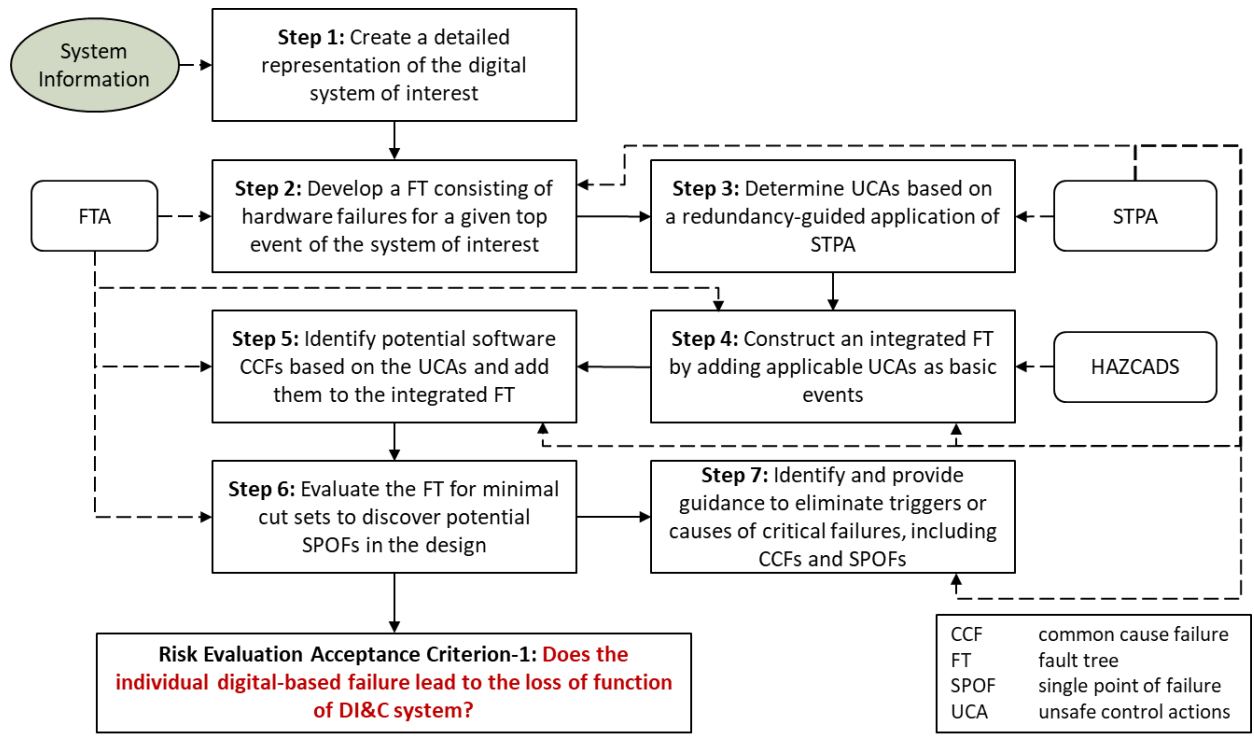

Figure 7. Workflow of the Redundant-guided System-theoretic Hazard Analysis (RESHA) in the PRADIC for the hazard analysis of DI&C systems (derived from [9]).

## 3.2 Case Studies

The RESHA method has been demonstrated based on a representative four-division digital RTS and ESFAS, which were modeled based on the DI&C design of Advanced Power Reactor 1400 MW (APR-1400) [18]. The functional logic of RTS and ESFAS are shown in Figure 8 and Figure 9 (derived from [10] and [11]). Both two DI&C systems have multilevel redundancy designs, where different levels of CCFs may occur because of different coupling mechanisms. For instance, there are 16 logic processors (LPs) in the local coincidence logical (LCL) racks that receive trip signals from bistable processors (BPs) and transmit the signal to next digital modules. There are four divisions, each division has two LCL racks, and each rack has two LPs. The LPs within the LCL racks require at least one-out-of-two (1oo2) BP signals per division and at least 2oo4 divisions to transmit a reactor trip signal. Therefore, three levels of CCFs may happen in LPs: (1) malfunction of all 16 LPs, (2) malfunction of 4 LPs in one division, and (3) malfunction of 2 LPs in one LCL rack. Here it is assumed that these LPs each have identical function and share the same features except for location of installation. CCFs are assumed to occur within a single location (e.g., division or rack) or across all divisions. Subsets of CCFs between divisions (e.g., A&B, B&C, and C&D) are not considered in RESHA because the main coupling factor in this case is the installation location. The final outputs from RESHA feed into following the PRADIC reliability analysis and provide component-level suggestions for system modifications (such as the elimination of CCFs that lead to potential SPOFs).

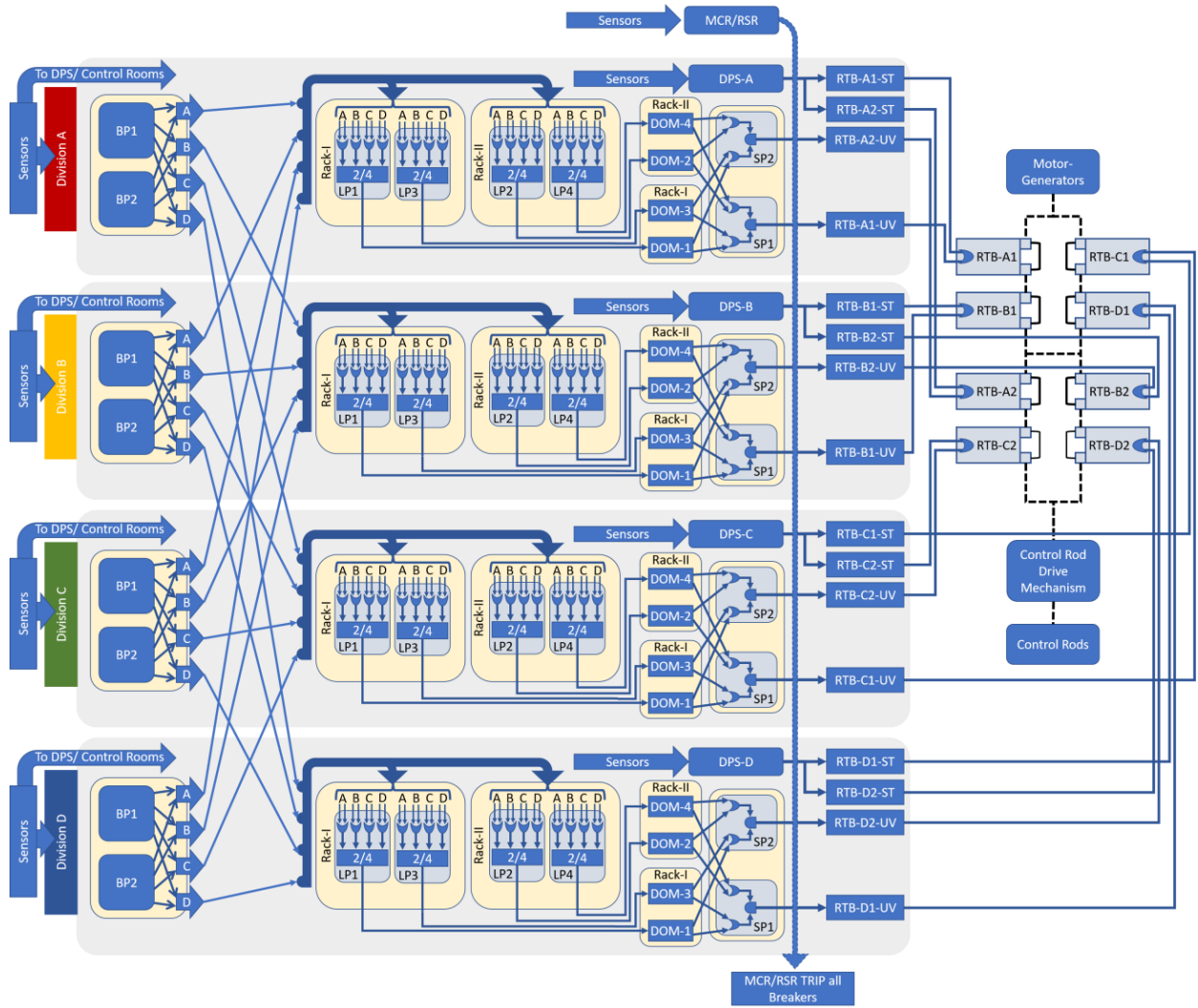

Figure 8. Functional logic of a representative four-division digital RTS (derived from [10]).

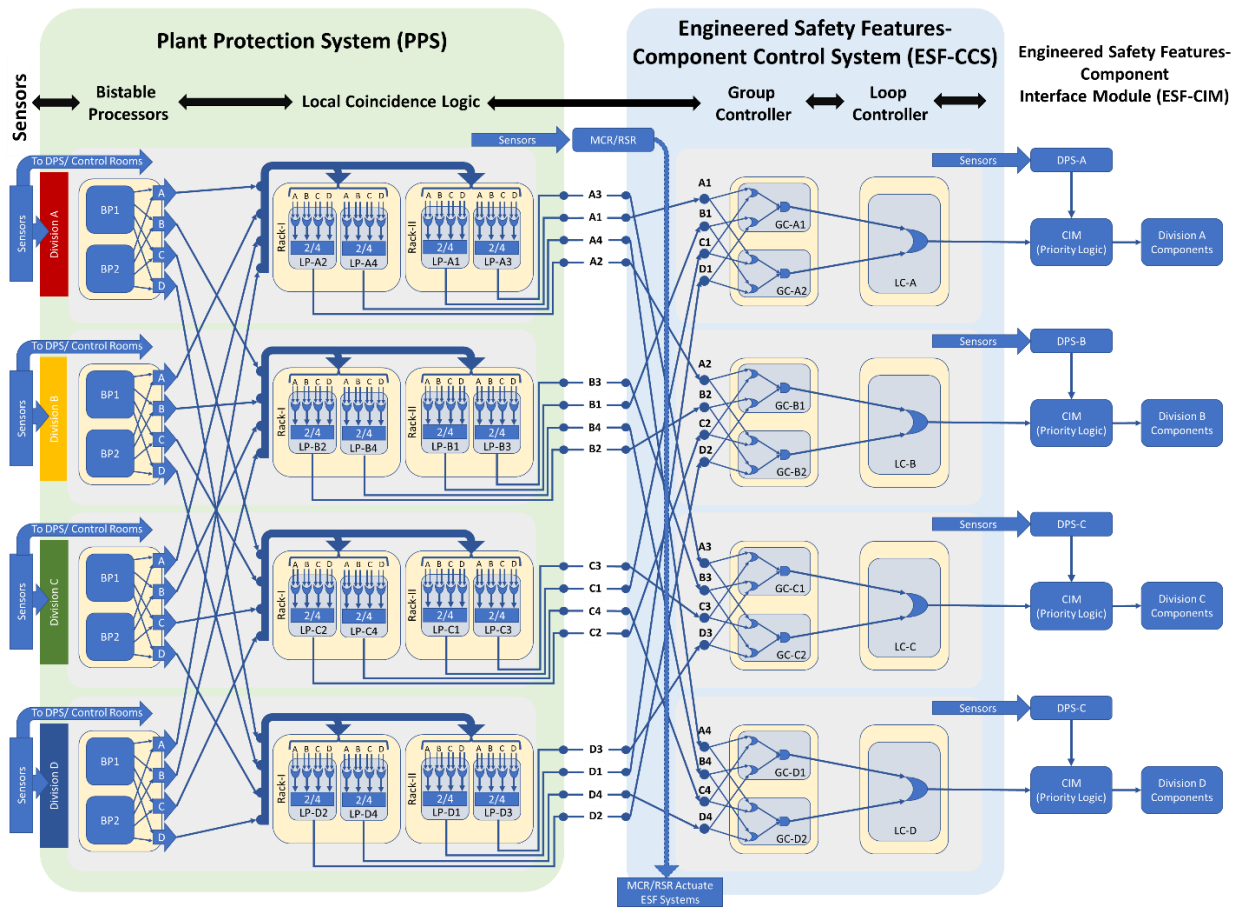

Figure 9. Functional logic of a representative four-division digital ESFAS (derived from [11]).

The Step 3 of RESHA method determines UCAs based on a redundancy-guided application of STPA, which develops a redundancy-guided multilayer control structure for the highly redundant DI&C systems. Figure 10 illustrates a redundancy-guided multilayer control structure for a digital ESFAS (derived from [11]). The top layer of redundancy is the division-level redundancy; four independent divisions can be used to actuate ESF components (i.e., high-pressure injection [HPI], low-pressure injection [LPI]). The functioning of each ESF component shown in Figure 9 is affected by a respective division. In each ESFAS division, two independent LCL racks receive and transmit actuation signals, which is the second layer of redundancy (i.e., unit-level redundancy). The third level of redundancy, the module level redundancy, consists of two LPs in each LCL rack. Therefore, three levels of CCFs in LP malfunction can be captured by RESHA: (1) LP CCF in all division, (2) LP CCF in one single division, and (3) LP CCF in one single LCL rack. Similarly, two levels of BP CCFs and group controller (GC) CCFs and one level of loop controller (LC) CCF can be identified in the four-division digital ESFAS-FT.

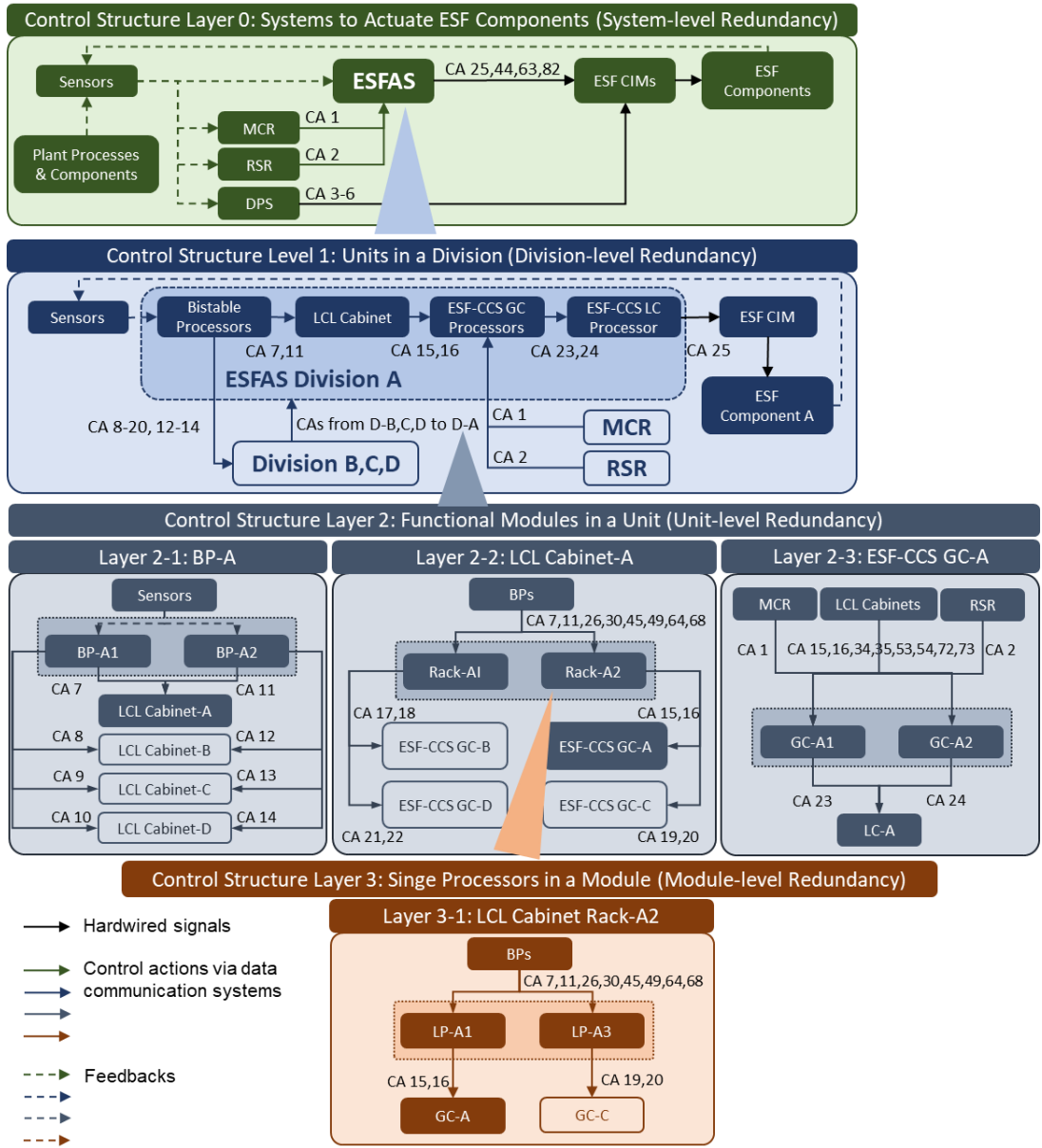

Figure 10. Redundancy-guided multilayer control structure for a digital ESFAS (derived from [11]).

The final outputs from RESHA feed into following the PRADIC reliability analysis and component-level risk evaluation that provides a guidance for system modifications (e.g., elimination of SPOFs, enhancement of reliability of specific components or diversity in designs). In a word, the PRADIC-RESHA provided a means to identify CCFs in digital-based Type II interactions and software of highly redundant HSSSR DI&C systems, by fully considering redundancy into the hazard analysis process. Part of the integrated FTs of the representative four-division digital RTS can be found in the Appendix A.

### 4. Multiscale Quantitative Reliability Analysis

The goal of the PRADIC reliability analysis is to estimate the DI&C system reliability by calculating the integrated FT of DI&C systems obtained in the hazard analysis, then provide inputs for following consequence analysis. For the reliability analysis of large-scale DI&C systems, the quantitative small-scale reliability analysis of software and Type II interactions in DI&C systems are also included in the PRADIC reliability analysis workflow.

## 4.1 Technical Approach

Figure 11 illustrates the framework of the PRADIC multiscale reliability analysis of DI&C system. The first step to any good reliability analysis for a DI&C system is the adequate collection and evaluation of design and requirement documentation. The target documents required, based on IEEE-guided software development lifecycle, include but are limited to, the software requirement specifications (SRS), the software design description (SDD), and the software test documentation (STD). These documents are necessary to determine if design and test failure data are available to conduct detailed and highly relevant reliability analysis. The SRS document highlights the specific stakeholder requirements for the design and lays out the functional and non-functional requirements and use cases. The SRS document can be thought of as high-level design. Completeness and consistency of the requirements specifications are subject to the experience and policy compliance of the development team and guides the rest of the project. The SDD document is the detailed design of the SRS document and provides explicit guidance on software and development procedures to achieve goals outlined in the SRS. Architectural and schematical diagrams are common aspects of the SDD but may also include interface, data, and procedural designs specifying how each functional and non-functional requirement is addressed. Lastly, the STD is a set of test documents that outline the design and test implementation plan as well as test results for the different aspects of the system. For example, included in the STD could be the master test plan and report. The master test plan provides the overall test planning strategy and management for the different levels of tests to be implemented. The master test report summarizes the results of all tests accomplished including assessment on the quality of tests and testing efforts, potentially anomalies detected by the test strategies, and whether the final designed system meets all specified functional and non-functional requirements. The overall adequacy of the system design dependent on the experience of the team and may lead to data rich and sparse scenarios. It is recognized that for many engineering software projects, from both experienced and inexperienced teams, the lack of adequate design documentation is prevalent potentially due to constraints on the project. A solution is provided for each case in terms of the ORCAS or BAHAMAS methodology.

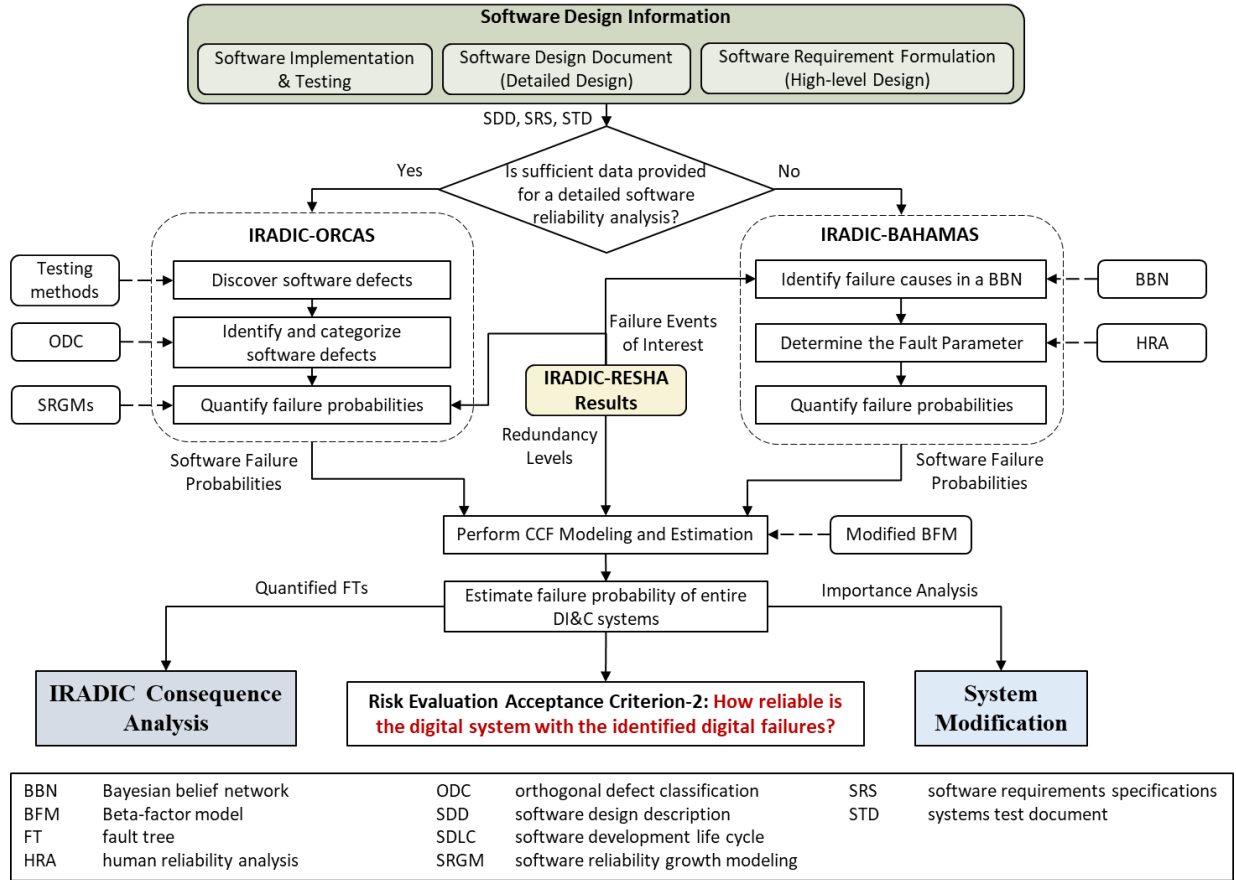

Figure 11. The framework of the PRADIC multiscale quantitative reliability analysis.

The next step is to estimate the failure probability of UCAs identified in the PRADIC hazard analysis. In PRADIC, two methods have been developed for different application conditions: BAHAMAS [14] for limited-data conditions and ORCAS for data-rich analysis. BAHAMAS was developed for the conditions with limited testing/operational data or for reliability estimations of software in early development stage. It can provide a rough estimation of failure probabilities to support the design of software and target DI&C systems even when data is very limited. Instead of relying on testing data, BAHAMAS assumes software failures can be traced to human errors in the software development life cycle (SDLC) and modeled with human reliability analysis (HRA). In BAHAMAS, a Bayesian belief network (BBN) is developed to provide a means of combining disparate causal factors and fault sources in the system, and HRA is applied to quantify the potential root human errors during SDLC. More technical information about BAMAHAS method can be found in [14]. In contrast to BAHAMAS, ORCAS applies a white box software invasive testing and modeling strategy to trace and identify the software defects that can potentially lead to software failures. The approach is a root-cause analysis methodology focused on comprehensive testing and follows the orthogonal defect classification (ODC) approach to determine testing sufficiency. ODC is also used to systematically classify the identified defects into specific software causality groups, thereby linking defects to potential software failure modes. The failure data collected from testing strategies is also combined with software reliability growth models and linear reliability models to quantify the software failure probability of specific UCAs. Currently, ORCAS is still under development, and recent demonstration will be published in a separate paper. The case study of software reliability analysis described in this paper was performed using BAHAMAS.

After the small-scale reliability analysis of software and Type II interactions, a modified beta-factor model (BFM) is applied for the modeling and estimation of CCFs, especially for software CCFs. Finally, when the basic events of integrated FT are calculated, the failure probability of entire DI&C system can be estimated using FTA tools.

### 4.2 Small-scale Software Reliability Analysis in a Data-limited Condition

This section describes how BAHAMAS is applied to quantify the software failure probability of a component (e.g., a UCA of BP) in a four-division RTS (in Figure 8) when design and testing data are very limited. According to [14], the principles assumptions for BAHAMAS are: (1) software failures can be traced to human errors in the SDLC and can be modeled with HRA methods; (2) in the absence of testing data, the reliability of a system can be predicted based on how its SDLC quality compares with existing systems of similar design and purpose. The BAMAHAS information flow is shown in Figure 12, which follows a six-step process: (1) identify a software failure of interest, (2) identify potential causes of the failure, (3) organize potential causes of the failure into a Bayesian network, (4) determine the fault parameter, (5) determine the probability for the failure of interest, and (6) evaluate the single and CCF probability [14].

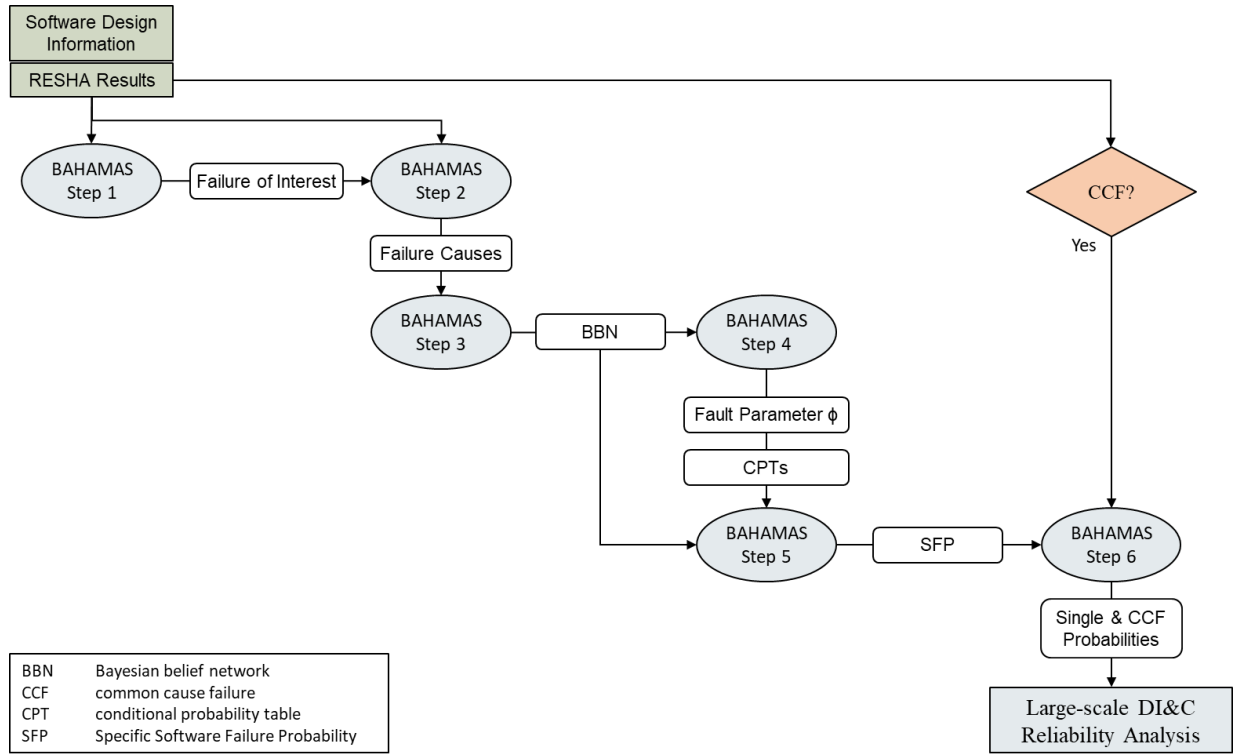

Figure 12. Information flow of PRADIC-BAHAMAS for the software reliability analysis of DI&C systems (derived from [14]).

One of the key outcomes of BAHAMAS is a BBN that consists of the root (primarily human) causes of failure for the controller culminating in a node that represents the probability of faults existing in system, as shown in Figure 13. Here only software-based faults are included in the network. Hardware-based component faults are separately incorporated into the RESHA integrated FT as hardware failures.

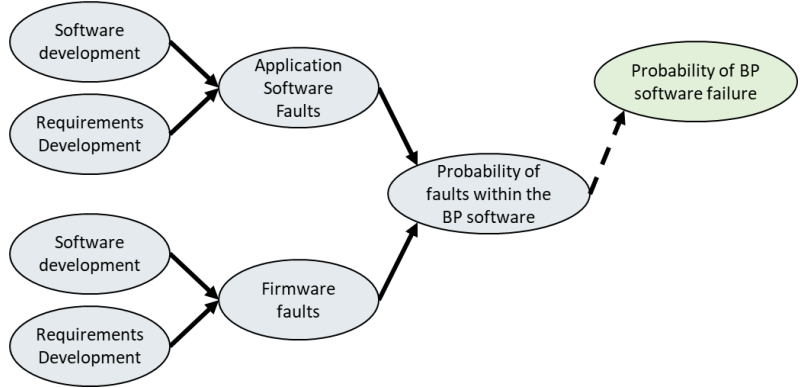

Figure 13. The BBN for a UCA of BPs of the four-division digital RTS based on current case study assumptions (derived from [14]).

BAHAMAS expresses specific software failure probability for a specific system by scaling the quality of its SDLC to that of existing designs in a linear relationship, as shown in Equation (1). Where $\phi$ represents the ratio of specific failure probability (SFP) to SDLC for a class of software systems.

$$\frac{SFP_{Specific}}{Specific\ Quality\ of\ SDLC} = \frac{SFP_{Generic}}{Generic\ Quality\ of\ SDLC} = \phi \tag{1}$$

The measure for the quality of the SDLC is given by the probability that there are (undetected) faults within the software. Equation (2) forms an expression for specific systems SFP as a function of phi and the probability of faults within the software.

$$SFP_{Specific} = \phi \cdot P(faults)_{Specific} \tag{2}$$

In this case, the SFP for the UCA of BP is calculated as $1.871E-4$, which is the sum of independent failures and CCFs of a UCA of BP. More details about this case study can be found in [14]. The next section will introduce how to model and estimate CCF probabilities based on the SFP.

### 4.3 Software Common Cause Failure Modeling and Estimation

#### 4.3.1 Method Description

A CCF is the result of the existence of two main factors—a failure cause and a coupling factor (or mechanism) [19] [20] [21]. The failure cause is the condition to which failure is attributed. The coupling factor (or coupling mechanism) creates the condition for a failure cause to affect multiple components thereby producing a CCF [19]. Examples of common coupling mechanisms given in NRUEG/CR-5485 include design, hardware, function, installation, maintenance, procedures, interfaces, locations, and environment. Any group of components which share similarities via coupling mechanisms may have vulnerability to CCF; a group of such components are considered a common cause component group (CCCG) [19]. The identification of coupling factors and CCCGs is an essential part of the CCF analysis. The typical process for CCF analysis is to assess a system, identify hazards, determine CCCGs, select CCF models, define model parameters, and evaluate the CCFs and their influences. Most of the preliminary steps are covered in the early portions of PRADIC. This section focuses on the quantification process starting from the identification of CCCGs for software failures.

Often CCF models attempt to simplify an analysis by assuming symmetry for components of the CCCG. For example, a CCCG may be created by assuming components are similar, and any subtle differences in coupling factors might be ignored. Nearly all CCF models rely on symmetry (the most notable exception being the BBN-based approaches) [22]. However, placing components into unique, coupling-factor-based CCCGs that account for the inconsistency between the groups may result in issues when relying on traditional techniques. Allowing a component to be part of multiple groups may lead to double counting of failure events or difficulties in quantification [13] [23].

In addition, most methods are designed to incorporate some form of operational data. A major challenge is the issue of limited data which has a direct influence on all methods and nearly guarantees the dependence on elicitation techniques. Therefore, rather than making special exceptions for conventional methods, other options were investigated. One method from 2012 was specifically developed for the multiple CCCG scenario and is based on a ratio approach like that of the beta-factor method (BFM) [24]. This method, the modified BFM, was designed specifically to allow for components to be members of multiple CCCGs. The model assumes the total failure probability/rate ($Q_t$) of a component is the summation of independent ($Q_I$) and dependent failures ($Q_D$). Equation (4) shows the total dependent failure consists of the contribution of each CCCG failure, where each CCCG is assigned a group beta ($\beta_w$). Each group beta represents the contribution of a single CCCG to the total failure probability. Equation (8) shows the independent failure probability in terms of each CCCG beta and total failure probability [24].

$$Total\ failure\ probability = Q_t = [independent\ failure] + [dependent\ failures] \quad (3)$$

$$Q_D = P(CCCG_1) + P(CCCG_2) + \cdots P(CCCG_w) \quad (4)$$

$$P(CCCG_w) = (\beta_w)Q_t \quad (5)$$

$$\beta_t = \sum_1^w (\beta_w) \quad (6)$$

$$Q_D = Q_t \sum_1^w (\beta_w) \quad (7)$$

$$Q_I = (1 - \beta_t)Q_t = \left[1 - \sum_1^w (\beta_w)\right] Q_t \quad (8)$$

Some advantages of this method include its ability to account for multiple CCCGs directly. It also avoids the potential for double counting that is present in conventional methods [13]. Additional advantages of this method are its ease of application and its ability to consider the coupling factors unique to each CCCG. PRADIC employs the modified BFM for the quantification of CCFs because it works directly for the multiple CCCG scenario. Subsequent discussions provide details for the quantification of the model parameters.

The next challenge is to define and estimate the beta factors of the modified BFM given a limited-data scenario while also providing consideration of the unique qualitative attributes for each CCCG. There are at least two methods presented in the literature which express the elicitation of the CCF parameters without the use or dependence on operational data. These two methods, both of which are called "partial beta methods," develop model parameters from qualitative attributes. The chief difference among them is how they find an overall beta; partial beta factor (PBF)-A employs an additive scheme, while PBF-M relies on a multiplicative scheme [25].The PBF-M may under predict model parameters, and so the additive scheme was selected [13]. PBF-A was developed by R. A. Humphreys and Rolls-Royce and Associates [26] and later became part of Unified Partial Method (UPM) [27]. The PBF-A method was founded on the question, "What attributes of a system reduce common cause failures?" [26] A collection attributes, called sub-factors, were selected which are known to contribute to the prevention of CCFs. The sub-factors are shown in Table 1. Each subfactor was weighted by reliability engineers for their importance. The methodology requires the analyst to assign a score (A, B, C, etc.) for each subfactor. An "E" indicates a component is well defended against CCFs (i.e., A= worst, E = best). Beta is then determined as a function of the assigned scores using Equation (9). The model was arranged such that the upper and lower limits for beta correspond with values reported in literature [26]. This is ensured by the subfactor weighting and the denominator given in Equation (9). The beta value determined by this method is intended to be used with the BFM.

Table 1. Beta-factor estimation table for hardware.

| Sub-factors | A | A+ | B | B+ | C | D | E |
|---|---|---|---|---|---|---|---|
| Redundancy (& Diversity) | 1800 | 882 | 433 | 212 | 104 | 25 | 6 |
| Separation | 2400 | | 577 | | 139 | 33 | 8 |
| Understanding | 1800 | | 433 | | 104 | 25 | 6 |
| Analysis | 1800 | | 433 | | 104 | 25 | 6 |
| MMI | 3000 | | 721 | | 173 | 42 | 10 |
| Safety Culture | 1500 | | 360 | | 87 | 21 | 5 |
| Control | 1800 | | 433 | | 104 | 25 | 6 |
| Tests | 1200 | | 288 | | 69 | 17 | 4 |
| Denominator for Equation (9), $d = 51000$ | | | | | | | |

Note: The values given for the table and denominator are slightly different than those given in the source material. The current table relied on an automated calculation based on Humphreys' original derivation. According to Humphreys, scoring A for all subfactor categories will result in 0.3 for the beta factor [27]. The current table provides 0.300 while the original provides 0.302. The difference is negligible, so this work employed the automated calculation for convenience.

$$\beta = \frac{\sum(Sub-factors)}{d} \quad (9)$$

The subfactor names alone are not sufficient to describe the details for assessing each actual subfactor; therefore, readers are advised to visit the original source material for making assessments. The advantage of this method is it is simple to apply and allows for a more structured determination of beta than simple judgments. The provided model allows for qualitative features to be considered in the quantification of CCFs. An approach for performing CCF analysis given the limited data and multiple CCCGs was developed by integrating modified BFM and PBF-A. The approach relies on the modified BFM to account for multiple CCCGs and PBF-A to define beta factors for each CCCG. The hybrid approach provides a means to overcome the limitations of conventional methods. A formalized process that relies on the modified BFM and PBF-A is shown in Figure 14. The primary sections' headings of Figure 14 come from descriptions of CCF modeling processes found in existing sources [19] [22]. The subsequent section will demonstrate this process as part of a case study.

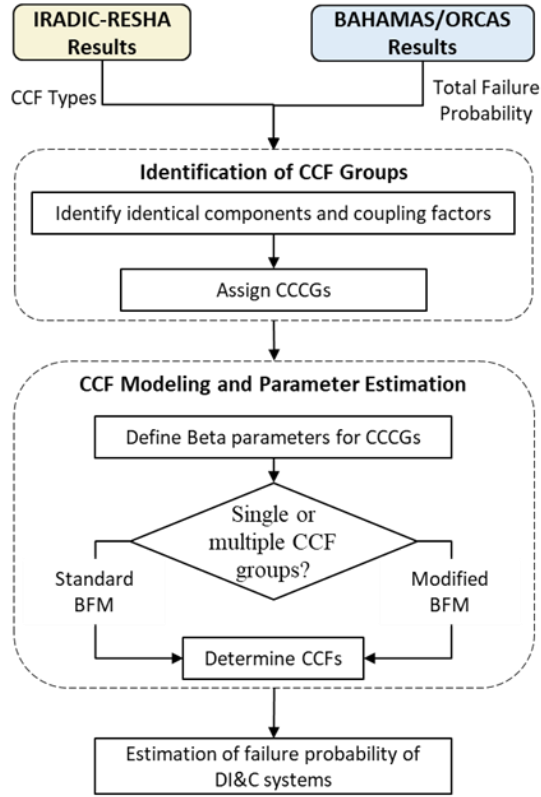

Figure 14. A CCF modeling flowchart developed for software CCF modeling and estimation in PRADIC reliability analysis.

### 4.3.2 Case Study

This case study describes the quantification of the CCFs found in the four-division digital RTS shown in Figure 8. The case study details the quantification process for BPs of the RTS. The following are the key assumptions of this study: (1) there is no diversity in the software, (2) all hardware components are identical (unless otherwise specified), (3) installation teams and maintenance teams are assumed identical for each CCCG, (4) each set of identical components that are part of the same CCCGs have the same total failure probabilities, and (5) software probability for the BP and LCL processors were quantified by BAHAMAS. They are assumed identical given limited details available to distinguish them.

Table 2 provides the list of components for which failure rates need to be quantified. The initial demonstration of BAHAMAS assumed a generic layout for these components, consisting of four parts: an input, output, central processing unit (CPU), and memory; each part was assumed to have software. These original assumptions have been modified to reflect new details for the RTS. In this work, the only components shown in Figure 8 to contain software are the BP and LCL processors, both of which are PLC platforms. The software is housed in the memory of each PLC processor. Thus, the evaluation with BAHAMAS followed the same format as given in the original publications [14] but modified to the assumptions (e.g., controller layout) of the current case study. Evaluating the software CCF values will follow the same approach described in Section 4.3.1 but with some modification for software-specific considerations.

Table 2. Total hardware and software failure probabilities for CCF case study.

| Components | Hardware Failure | Total Hardware Failure Probability | Software Failure | Total Software Failure Probability |
|---|---|---|---|---|
| BPs | YES | 4.00E-5 | YES | 1.871E-4 |
| LCL Processors | YES | 6.48E-5 | YES | 1.871E-4 |
| Digital Output Modules | YES | 1.64E-5 | N/A | N/A |
| Selective Relay | YES | 6.20E-6 | N/A | N/A |
| RTB-UV device | YES | 1.70E-3 | N/A | N/A |
| RTB-Shunt device | YES | 1.20E-4 | N/A | N/A |
| RTB RTSS1 | YES | 4.50E-5 | N/A | N/A |
| RTB RTSS2 | YES | 4.50E-5 | N/A | N/A |
| All hardware values came from [28]. Software probability for the BP and LCL processors were quantified by BAHAMAS. They are assumed identical given limited details available to distinguish them. | | | | |

The PBF-A was designed for physical components and systems, including electrical systems, with only minimal consideration of software. In contrast, our current work requires greater emphasis on software attributes. The case study contains highly redundant software components with minimal redundancy. Given the lack of diversity, it is anticipated that CCFs should represent a significant portion of a component's total software failure. In addition, the PBF-A was clearly developed to assess hardware CCFs. To account for the needs of the case study, a preliminary solution was developed by modifying the PBF-A in two ways: (1) the model was adjusted to provide a range of beta values (i.e., 0.001–0.999), allowing for greater applicability to low-diversity software systems; (2) the subfactor weights were changed to emphasize software-centric features. The adjusted model emphasizes the introduction of software faults and coupling factors by placing greater weight on those defenses that pertain to human interaction and diversity of software. Table 3 shows the modified PBF-A beta-factor estimation table. It, along with Table 1, are used for the define the beta factors for software and hardware failures respectively.

Table 3. Beta-factor estimation table for software.

| Sub-factors | A | A+ | B | B+ | C | D | E |
|---|---|---|---|---|---|---|---|
| Redundancy (& Diversity) | 23976 | 10112 | 4265 | 1799 | 759 | 135 | 24 |
| Separation | 23976 | | 4265 | | 759 | 135 | 24 |
| Understanding | 7992 | | 1422 | | 253 | 45 | 8 |
| Analysis | 7992 | | 1422 | | 253 | 45 | 8 |
| MMI | 11988 | | 2132 | | 379 | 67 | 12 |
| Safety Culture | 6993 | | 1244 | | 221 | 39 | 7 |
| Control | 4995 | | 888 | | 158 | 28 | 5 |
| Tests | 11988 | | 2132 | | 379 | 67 | 12 |
| Denominator for Equation (9), $d = 100000$ | | | | | | | |

Following the process shown in Figure 14, CCCGs can be assigned after identifying identical components and their coupling factors. There are eight identical BPs in the RTS, two per division. They each have identical function and are assumed to share the same features except for location of installation. All BPs share identical coupling factors except for location resulting in two CCCGs. One CCCG is based on shared function, hardware, software, and manufacturer. The second CCCG considers location. Table 4 shows the CCCGs for the BPs.

Table 4. CCCGs for the BPs.

| CCCGs | | Coupling Factors |
|---|---|---|
| 1 | All BPs | Function, Hardware, Software, & Manufacturer |
| 2 | Division A: BP1, BP2 | Division A |
| 3 | Division B: BP1, BP2 | Division B |
| 4 | Division C: BP1, BP2 | Division C |
| 5 | Division D: BP1, BP2 | Division D |

The next step from Figure 14 is to define the beta-factor parameters to be used for CCF modeling. The PBF-A method evaluates the defense categories which pertain to the prevention of CCFs. Each CCCG receives a score for each subfactor category. Table 5 shows the subfactor scores applied to the BPs of CCCG 1 and the calculation for beta based on Equation (9). The BPs for CCCGs 2–5 share the same qualitative features and receive beta-factor scores of 0.123 and 0.568 for hardware and software, respectively.

Table 5. Subfactor scores for BPs CCCG 1 (all BPs CCF).

| Sub-factors | Hardware | | Software | |
|---|---|---|---|---|
| Redundancy (& Diversity) | B+ | 212 | A | 23976 |
| Separation | E | 8 | A+ | 10112 |
| Understanding | A | 1800 | A | 7992 |
| Analysis | D | 25 | D | 45 |
| MMI | C | 173 | C | 379 |
| Safety Culture | E | 5 | E | 7 |
| Control | D | 25 | D | 28 |
| Tests | C | 69 | C | 379 |
| Beta for the CCCG | $\beta_{HD1} = 0.045$ | | $\beta_{SW1} = 0.429$ | |

The next step from the CCF modeling flowchart is to determine the CCFs. The BPs have multiple CCCGs; therefore, the modified BFM is used. Division A BP1 is found in two groups, CCCG1 and CCCG2, as shown in Table 6. Equations (3)–(8) are used to find the independent and dependent failures of the BPs.

The results of the CCF analysis are shown in Table 6 and Table 7. Note that RACK, DIVISION, and ALL, correspond to the CCCG categories, while INDIVIDUAL corresponds to individual component failure. The CCCG ALL contains all the identical components within the system of interest. The given CCCG categories are not shared by all components; hence, there are no RACK CCCGs for the RTBs.

Table 6. Hardware failure probability for RTS components.

| Component | INDIVIDUAL | RACK | DIVISION | ALL | Total |
|---|---|---|---|---|---|
| BPs | 4.000E-05 | N/A | 5.943E-06 | 2.187E-06 | 4.813E-05 |
| LCL Processors | 6.480E-05 | 1.076E-05 | 7.647E-06 | 3.961E-06 | 8.717E-05 |
| DOMs | 1.640E-05 | 1.706E-06 | 1.015E-06 | 1.983E-07 | 1.932E-05 |
| Selective Relay | 6.200E-06 | N/A | 6.073E-07 | 7.059E-08 | 6.878E-06 |
| RTB-UV device | 1.700E-03 | N/A | N/A | 1.763E-05 | 1.718E-03 |
| RTB-Shunt device | 1.200E-04 | N/A | N/A | 1.244E-06 | 1.212E-04 |
| RTB RTSS1 | 4.500E-05 | N/A | N/A | 1.944E-06 | 4.694E-05 |
| RTB RTSS2 | 4.500E-05 | N/A | N/A | 1.944E-06 | 4.694E-05 |

Table 7. Software failures probability for RTS components.

| Component | INDIVIDUAL | RACK | DIVISION | ALL | Total |
|---|---|---|---|---|---|

| BPs | 5.591E-07 | N/A | 1.062E-04 | 8.030E-05 | 1.871E-04 |
| LCL Processors | 8.086E-05 | N/A | N/A | 1.062E-04 | 1.871E-04 |

Regarding the results, there is a difference between the software and hardware CCCGs of the LCL processors. The hardware CCCG for LCL processor is separated by location just like the BPs. However, the potential for DIVISION and RACK level CCFs are precluded from consideration because there is nothing to distinguish them from the CCCGs representing all LCL processors; according to the case study, each LCL processor has the same software and receives the same inputs. By contrast, the BPs have the potential for input variation amongst divisions. Thus, the BPs have DIVISON level software CCCGs, but the LCL processors do not.

### 4.4 Large-scale DI&C System Reliability Analysis

By assigning the failure probabilities into the integrated FTs of four-division digital RTS and ESFAS, the failure probabilities of these two systems can be calculated using the INL-developed PRA tool SAPHIRE [29].

#### 4.4.1 Integrated Fault Tree for Digital RTS

The main logic of this integrated RTS-FT is displayed in Figure 17. This FT was quantified with SAPHIRE 8 using a truncation level of 1E-12; RTS failure probability is 1.270E-6 with 13 cut sets. Table 8 lists part of these cut sets with significant contributions. Mechanical CCF of rod control cluster assembly (RCCA) is the main contributor (>95% of total); the software CCFs do not significantly affect the reliability of digital RTS because of the highly redundant design and high reliability of digital components.

Table 8. Cut sets for the improved RTS-FT.

| FT Name | # | Probability | Total % | Cut Sets |
|---|---|---|---|---|
| Integrated RTS-FT | 1 | 1.210E-6 | 95.31 | RPS-ROD-CF-RCCAS |
|  | 2 | 2.052E-8 | 1.62 | RPS-CCP-TM-CHA, RPS-TXX-CF-4OF6, RPS-XHE-XE-NSIGNL |
|  | 3 | 1.944E-8 | 1.53 | RPS-XHE-XE-SIGNL, RTB-SYS-2-HD-CCF |
|  | 4 | 1.944E-8 | 1.53 | RPS-XHE-XE-SIGNL, RTB-SYS-1-HD-CCF |
|  | Total | 1.270E-6 | 100 | - |

#### 4.4.2 Integrated Fault Tree for Digital ESFAS

This FT was also quantified with SAPHIRE 8 using a truncation level of 1E-12; ESFAS failure probability is 2.095E-5 with one cut set. Hardware CCF of ESF-component interface modules (ESF-CIMs) is the main contributor, and the software CCFs do not significantly affect the reliability of digital ESFAS because of the high-redundant design and high reliability of digital systems.

### 5. Consequence Analysis

This section documents the consequence analysis of a generic PWR SAPHIRE model with integrated digital RTS and ESFAS FTs. Section 5.1 describes the generic PWR SAPHIRE model, and the scenario to be analyzed is introduced, as well as the original ET models for these scenarios including a FT for the failure of an analog RTS and one CCF basic event for the ESFAS failure. Section 5.2 compares the original FTs for analog RTS and ESFAS and the new integrated FTs for digital RTS and ESFAS. Results for consequence analysis of these selected ET models are discussed in Section 5.3. In previous work, a consequence analysis was performed by focusing on IE-TRANS with RTS-FT; other ETs or ESFAS-FT was not considered [9].

## 5.1 Introduction of INL Generic PWR SAPHIRE Model

A generic internal events PRA model was developed at INL using SAPHIRE 8 for a typical PWR plant for the accident scenario analysis with various initiating events, which has been applied for different purposes including plant-level scenario-based risk analysis for enhanced resilient plant (ERP) during station black-out (SBO) and loss-of-coolant accident (LOCA) [30], risk-informed analysis for an ERP with Accident Tolerant Fuel (ATF), optimal use of Diverse and Flexible Coping Strategy (FLEX), and new passive cooling systems [31] [32]. There are 24 ETs included in this generic PWR SAPHIRE model. This work applies SAPHIRE 8 for the FT development of DI&C system and combines these integrated FTs with the existing generic PWR ET models. The consequence analysis of DI&C failures documented in this paper covers the following accident scenarios: INT-TRANS (initiating event - general plant transient) with ATWS (anticipated transient without scram), LOSC (loss-of-seal cooling), SBLOCA (small-break loss-of-coolant accident), and MBLOCA (medium-break LOCA). These five accident ETs are respectively shown from Figure 21 to Figure 25. INT-TRANS is selected here for the DI&C consequence analysis because it shows relatively significant impacts of digital failures to key plant responses. RTS and ESFAS failures are treated as initiating events or important basic events included in cut sets that have significant contributions to change of CDF (ΔCDF).

## 5.2 Original and Improved Integrated Fault Trees for HSSSR DI&C Systems

### 5.2.1 Original Fault Tree for Reactor Trip System

The original RTS-FT in the generic PWR SAPHIRE model has identified analog/hardware failures in detail. The main logic of original RTS-FT is shown in Figure 15. A two-train analog RTS was modeled; the main failure modes include electric failures, CCF of RCCA fail to drop, contribution of seismic events, operator errors, and RTS failures during test and maintenance. This FT was quantified with SAPHIRE 8 using a truncation level of 1E-12; RTS failure probability is 4.288E-6 with five cut sets. Results indicate hardware CCFs are the main concerns for the failure analog safety-related redundant I&C systems. Compared with the original RTS-FT, the total failure probability of integrated four-division RTS-FT is reduced about 50%.

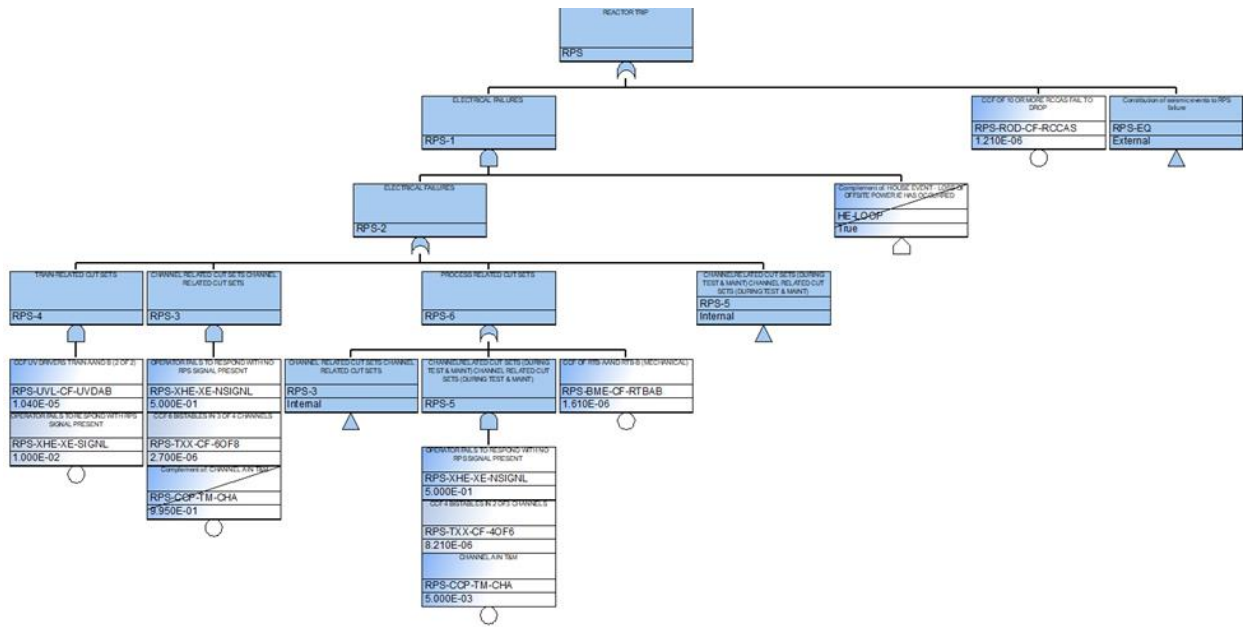

Figure 15. Main FT of original RTS-FT in the generic PWR SAPHIRE model.

### 5.2.2 Original Fault Tree for Engineered Safety Features Actuation System

In the original generic PWR SAPHIRE model, ESFAS failure is presented using a CCF of ESF actuation signal in both Train A and B, named ESF-VCF-CF-TRNAB with a probability as 6.420E-4. These CCF basic events are used in the FTs of several top events in IE-TRANS scenarios including AFW (representing failure of auxiliary feedwater), AFW-ATWS (representing failure of auxiliary feedwater for ATWS scenarios), HPI (representing failure of high-pressure injection), and LPI (representing failure of low-pressure injection). It should be noted another basic event is used to represent "operator fails to manually initiate safety features," which will be replaced by an integrated FT representing failure of digital human system interface and operator errors in future work.

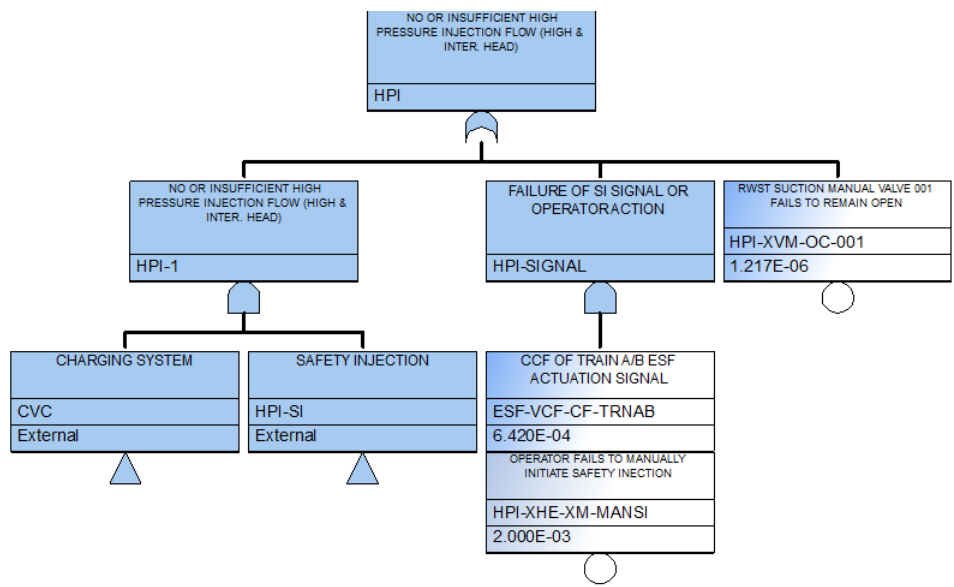

Figure 16. Main FT of HPI failure in the generic PWR SAPHIRE model where CCF of analog ESFAS is considered.

Compared with the original ESFAS failure, the total failure probability of ESFAS is significantly reduced. Table 9 compares the number of cut sets and failure probabilities of two FTs.

Table 9. Cut sets of the improved ESFAS-FT.

| FT Name | Probability | Total % | # of Cut Sets |
|---|---|---|---|
| **Improved ESFAS-FT** | 2.095E-5 | 100 | 1 |
| **Original ESFAS failure** | 6.420E-4 | 100 | 1 |

Accordingly, the impact of ESFAS failure to the actuations of safety features were estimated by solving the FTs of AFW, AFW-ATWS, HPI, and LPI; results are listed and compared in Table 10. All the failure probabilities of these safety features have been reduced due to the decrease of ESFAS failure probability.

Table 10. Comparison of the top events with original ESFAS-CCF basic event and improved ESFAS-FT.

| Top Event | Probability | | # of Cut Sets | |
|---|---|---|---|---|
| | Original | New | Original | New |
| **Failure of AFW** | 1.487E-5 | 1.231E-5 | 1539 | 1539 |
| **Failure of AFT-ATWS** | 2.367E-4 | 2.342E-4 | 906 | 906 |
| **Failure of HPI** | 1.104E-5 | 9.756E-6 | 1163 | 1163 |
| **Failure of LPI** | 8.416E-4 | 2.018E-4 | 1567 | 1567 |

### 5.3 Accident Scenario Analysis for General Plant Transient

In this section, IE-TRANS, IE-SBLOCA, and IE-SBLOCA ET tree quantification results are discussed. As IE-ATWS and IE-LOSC are sub-ETs of INT-TRANS developed in the generic PWR SAPHRIE model, they will be discussed in the section on INT-TRANS.

#### 5.3.1 INT-TRANS

The generic PRA model representing general plant transient as INT-TRANS was shown in Figure 21. The IE-TRANS ET was quantified with SAPHIRE 8 using a truncation level of 1E-12. Table 11 compares the quantified CDF with original and new FTs. The original total IE-TRANS CDF is 1.073E-6/year and half-reduced to 5.769E-7/year with the new FTs. There are 16 non-zero CDF sequences out of a total of 145 INT-TRANS accident sequences (i.e., the sequence end state is core damage).

INT-TRANS:21-16 from ATWS scenarios is one of the most risk-significant sequences with a CDF reduced from 5.388E-7/year to 1.595E-7/year and contributes 27.6% of the CDF of improved INT-TRANS. In this sequence, RTS fails to trip the reactor; primary and secondary side depressurizations are not successful because safety relief values are closed. Core damage occurs as long-term cooling cannot be established.

INT-TRANS:20 from TRANS scenarios is another risk-significant sequence with a CDF reduced from 3.895E-7/year to 3.262E-7/year, contributing 56.5% of the CDF of improved INT-TRANS. In this sequence, RTS successfully trips the reactor, core damage occurs as neither auxiliary feedwater is not available nor HPI could provide makeup water to the reactor coolant system.

Table 11. Comparison of INT-TRANS ET quantification results.

| Sequence | CDF (per reactor year) | | | # of Cut Sets (Truncated by 1E-12) | |
|---|---|---|---|---|---|
| | Original ET | Improved ET | Δ CDF/ Original CDF | Original ET | Improved ET |
| INT-TRANS:21-16 | 5.388E-07 | 1.595E-07 | -70.40% | 51 | 38 |
| INT-TRANS:20 | 3.895E-07 | 3.262E-07 | -16.25% | 1060 | 1041 |
| INT-TRANS:21-14 | 7.262E-08 | 2.149E-08 | -70.41% | 49 | 18 |
| INT-TRANS:02-02-09 | 5.830E-08 | 5.830E-08 | 0 | 1248 | 1248 |
| INT-TRANS:19 | 8.132E-09 | 6.692E-09 | -17.71% | 282 | 236 |
| INT-TRANS:02-03-09 | 2.731E-09 | 2.731E-09 | 0 | 387 | 387 |
| INT-TRANS:02-02-10 | 9.546E-10 | 9.546E-10 | 0 | 168 | 168 |
| INT-TRANS:21-15 | 7.568E-10 | 2.124E-10 | -71.93% | 102 | 29 |
| INT-TRANS:02-04-10 | 5.865E-10 | 5.865E-10 | 0 | 142 | 142 |
| INT-TRANS:02-14-10 | 1.994E-10 | 1.994E-10 | 0 | 81 | 81 |
| INT-TRANS:02-03-10 | 7.653E-12 | 7.653E-12 | 0 | 4 | 4 |
| INT-TRANS:02-09-09 | 7.558E-12 | 7.558E-12 | 0 | 4 | 4 |
| INT-TRANS:02-06-09 | 7.558E-12 | 7.558E-12 | 0 | 4 | 4 |
| INT-TRANS:02-08-09 | 7.558E-12 | 7.558E-12 | 0 | 4 | 4 |
| INT-TRANS:02-07-09 | 2.287E-12 | 2.287E-12 | 0 | 2 | 2 |
| INT-TRANS:02-10-09 | 2.287E-12 | 2.287E-12 | 0 | 2 | 2 |
| Total | 1.073E-6 | 5.769E-7 | | 3590 | 3408 |

### 5.3.2 INT-SLOCA

The generic PRA model representing small-break LOCA as INT-SLOCA was shown in Figure 24. The INT-SLOCA ET was quantified with SAPHIRE 8 using a truncation level of 1E-12. Table 12 compares the quantified CDF with original and new FTs. The original total INT-SLOCA CDF is 7.784E-8/year and reduced to 7.509E-8/year with the new FTs. There are seven non-zero CDF sequences out of a total of 10 INT-SLOCA accident sequences (i.e., the sequence end state is core damage).

INT-SLOCA:03 is the most risk-significant sequences with a CDF of 6.433E-8/year and contributes 85.7% of the CDF of improved INT-SLOCA. In this sequence, RTS successfully trips the reactor, and the auxiliary feedwater and HPI are available. Core damage still occurs, as long-term low-pressure cooling cannot be established.

Table 12. Comparison of INT-SLOCA ET quantification results.

| Sequence (with CDF > 1E-9) | CDF (per reactor year) | | | # of Cut Sets (Truncated by 1E-12) | |
|---|---|---|---|---|---|
| | Original ET | Improved ET | Δ CDF/ Original CDF | Original ET | Improved ET |
| INT-SLOCA:03 | 6.433E-08 | 6.433E-08 | 0 | 564 | 564 |
| INT-SLOCA:05 | 2.867E-09 | 2.867E-09 | 0 | 97 | 97 |
| INT-SLOCA:09 | 7.386E-09 | 6.619E-09 | - 10.38% | 142 | 142 |
| Total | 7.784E-8 | 7.509E-8 | - 3.53% | 838 | 837 |

### 5.3.3 INT-MLOCA

The generic PRA model representing medium-break LOCA as INT-MLOCA was shown in Figure 25. The INT-MLOCA ET was quantified with SAPHIRE 8 using a truncation level of 1E-12. Table 13 compares the quantified CDF with original and new FTs. The original total INT-MLOCA CDF is 6.279E-7/year and reduced to 4.984E-7/year with the new FTs. There are eight non-zero CDF sequences out of a total of nine INT-MLOCA accident sequences (i.e., the sequence end state is core damage).

INT-MLOCA:03 is the most risk-significant sequences with a CDF of 4.917E-7/year and contributes 98.7% of the CDF of improved INT-MLOCA. In this sequence, RTS successfully trips the reactor, and the auxiliary feedwater and HPI are available. Core damage still occurs, as long-term low-pressure cooling cannot be established.

The most significant CDF reduction appears in INT-MLOCA:10 from 1.305E-07 to 2.567E-09, which contributes to about 99% of Δ CDF. In this sequence, RTS successfully trips the reactor, and the auxiliary feedwater is available. Core damage still occurs as neither HPI nor LPI can be established for long-term cooling. According to the change of failure probability of LPI shown in Table 10, LPI availability was highly increased by adding a FT of a more reliable four-division digital ESFAS instead of using a conservative CCF. Improved ET models can provide a more accurate prediction for ESFAS failure and relevant sequences.

Table 13. Comparison of INT-MLOCA ET quantification results.

| Sequence (with CDF > 1E-9) | CDF (per reactor year) | | | # of Cut Sets (Truncated by 1E-12) | |
|---|---|---|---|---|---|
| | Original ET | Improved ET | Δ CDF/ Original CDF | Original ET | Improved ET |
| INT-MLOCA:03 | 4.917E-07 | 4.917E-07 | 0 | 722 | 722 |
| INT-MLOCA:05 | 1.870E-09 | 1.870E-09 | 0 | 47 | 47 |
| INT-MLOCA:07 | 8.999E-11 | 8.999E-11 | 0 | 26 | 26 |
| INT-MLOCA:09 | 1.866E-09 | 1.866E-09 | 0 | 192 | 192 |
| INT-MLOCA:10 | 1.305E-07 | 2.567E-09 | -98.12% | 47 | 47 |
| INT-MLOCA:11 | 5.206E-10 | 8.654E-12 | -98.34% | 3 | 3 |
| INT-MLOCA:12 | 5.320E-10 | 3.510E-12 | -99.34% | 8 | 2 |
| INT-MLOCA:14 | 8.576E-10 | 2.539E-10 | -70.39% | 5 | 4 |
| Total | 6.279E-7 | 4.984E-7 | -20.62% | 1050 | 1043 |

### 5.4 Summary of Consequence Analysis

To compare the changes of CDF after adding integrated FTs of digital RTS and ESFAS to the generic PWR ET models; in this chapter, consequence analysis has been performed based on INT-TRANS and relevant accident scenarios. Integrated FTs of RTS and ESFAS include both software and hardware failures, particularly CCF, that may occur in a four-division digital RTS and a four-division digital ESFAS. Results show the CDF of INT-TRANS accident scenarios are reduced significantly.
The original generic PWR SAPHIRE model has 24 event trees; most of them use a top event for the failure of a two-train RTS. Its detailed FT consists of hardware failures, operator errors, and external hazards. ESFAS failure is modeled using a basic event as CCF of ESF actuation signal, which is embedded in the FTs of relevant safety features including auxiliary feedwater, HPI, and LPI. This CCF basic event is included in cut sets that have significant contributions to CDF. This generic PWR SAPHIRE model represents the conditions of existing U.S. NPPs with traditional analog HSSSR DI&C systems.

By adding the integrated FTs of four-division digital RTS and ESFAS into the PRA models, the safety margin obtained from the plant digitalization on HSSSR DI&C systems are quantitatively estimated. For example, results show RTS failure probability is half-reduced from 4.288E-6 to 1.270E-6; LPI failure probability greatly decreases from 8.416E-4 to 2.095E-4 due to the improvement of ESFAS-FT. This explains the significant reduction of CDF in these analyzed accident scenarios. Plant modernization including the improvement of HSSSR DI&C systems such as RTS and ESFAS will benefit plant safety by providing more safety margins to accident management.

The numbers of cut sets are also reduced due to the improved design from two-train analog I&C systems to a four-division DI&C systems. As the complexity of the system increases, the number of failure combination should also increase. However, with the improved design, the cut-set probabilities are reduced and truncated below the 1E-12 threshold.

## 6. Conclusions

This paper documents the research activities that quantitatively evaluate CCFs (especially software CCFs) in HSSSR DI&C systems (e.g., four-division digital RTS and ESFAS) in NPPs using the PRADIC. The PRADIC has been developed, demonstrated, and improved for the design of HSSSR DI&C systems with multilayer software CCFs, human interactions with these systems, and plant responses. This technology complements other approaches being developed for deploying DI&C technologies and emphasizes risk-informed approaches used to facilitate the adoption and licensing of HSSSR DI&C systems. Currently, only qualitative assessment is required for evaluating design attributes and quality measures of DI&C systems because there is no appropriate approach for performing quantitative assessment. To deal with the technical issues in addressing potential software CCF issues in HSSSR DI&C systems of NPPs, the PRADIC provides:

1. An integrated and best-estimate, risk-informed capability to address new technical digital issues quantitatively, accurately, and efficiently in plant modernization progress, such as software CCFs in HSSSR DI&C systems of NPPs
2. A common and modularized platform for I&C designers, software developers, plant engineers and risk analysts to efficiently prevent and mitigate risk by identifying crucial failure modes and system vulnerabilities, quantifying DI&C system reliability, and evaluating the consequences of digital failures on the plant responses
3. A technical basis and risk-informed insights to assist NRC and industry in formalizing relevant licensing processes relevant to CCF issues in HSSSR DI&C systems
4. An integrated risk-informed tool for vendors and utilities to meet the regulatory requirements and optimize the D3 applications in the early design stage of digital HSSSR systems.

As a tool for quantitative risk-informed analysis, the PRADIC aims to be practical, flexible, and grounded for different conditions of available, relevant and, appropriately evaluated (ARAE) data. By following a modularized data-adaptive flowgraph, the PRADIC can work with very limited data and refined data to achieve a method-data consistency. As data gets refined, the PRADIC produces higher fidelity output. The PRADIC also provides evidence to inversely inform the importance level of each ARAE data (which is necessary, which is optional) for safety assurance. The work in near future includes the improvement of CCF modeling for software failures, development of reliability analysis method (i.e., ORCAS) for data-rich conditions, relevant validation and uncertainty quantification of these quantitative methods, and adjust the PRADIC on the risk assessment and design optimization of AI-guided advanced control systems.

## 7. Acknowledgments

This submitted manuscript was authored by a contractor of the U.S. Government under DOE Contract No. DE-AC07-05ID14517. Accordingly, the U.S. Government retains and the publisher, by accepting the article for publication, acknowledges that the U.S. Government retains a nonexclusive, paid-up, irrevocable, worldwide license to publish or reproduce the published form of this manuscript, or allow others to do so, for U.S. Government purposes. This information was prepared as an account of work sponsored by an agency of the U.S. Government. Neither the U.S. Government nor any agency thereof, nor any of their employees, makes any warranty, express or implied, or assumes any legal liability or responsibility for the accuracy, completeness, or usefulness of any information, apparatus, product, or process disclosed, or represents that its use would not infringe privately owned rights. References herein to any specific commercial product, process, or service by trade name, trademark, manufacturer, or otherwise, does not necessarily constitute or imply its endorsement, recommendation, or favoring by the U.S. Government or any agency thereof. The views and opinions of authors expressed herein do not necessarily state or reflect those of the U.S. Government or any agency thereof.

The research activities and achievements documented in this paper were funded by the United States Department of Energy's Light Water Reactor Sustainability Program, Risk Informed Systems Analysis (RISA) Pathway. The authors would like to recognize the technical supports from Kenneth D. Thomas, Zhegang Ma, Ronald L. Boring, Sai Zhang, Thomas A. Ulrich, Jeffrey C. Joe, Congjian Wang, and Linyu Lin at Idaho National Laboratory. The authors also thank Rebecca N. Ritter at Idaho National Laboratory for technical editing and formatting of this paper. The authors would also like to acknowledge our collaborators from universities, Nam Dinh at North Carolina State University, Heng Ban at University of Pittsburgh, and Hyun Gook Kang at Rensselaer Polytechnic Institute, for their valuable comments in methodology development, and demonstration.

**Appendix A:** Part of Integrated Fault Tree for a Four-division RTS (derived from [9]).

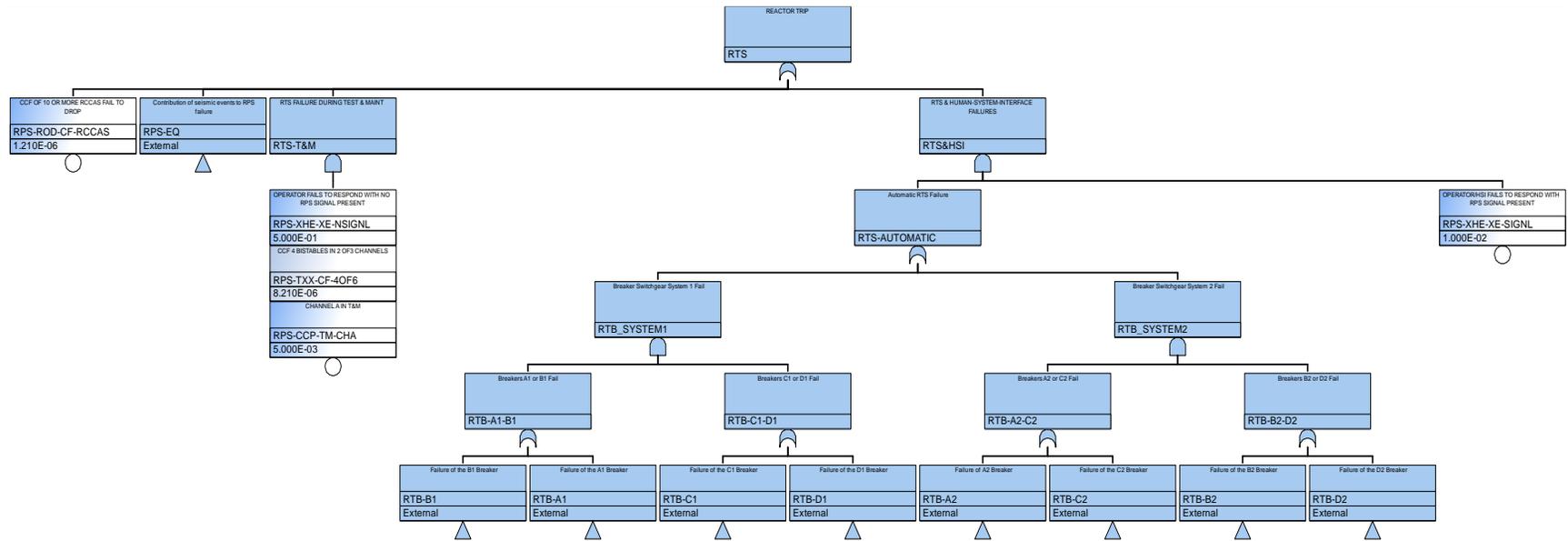

Figure 17. Main FT of integrated RTS-FT using PRADIC.

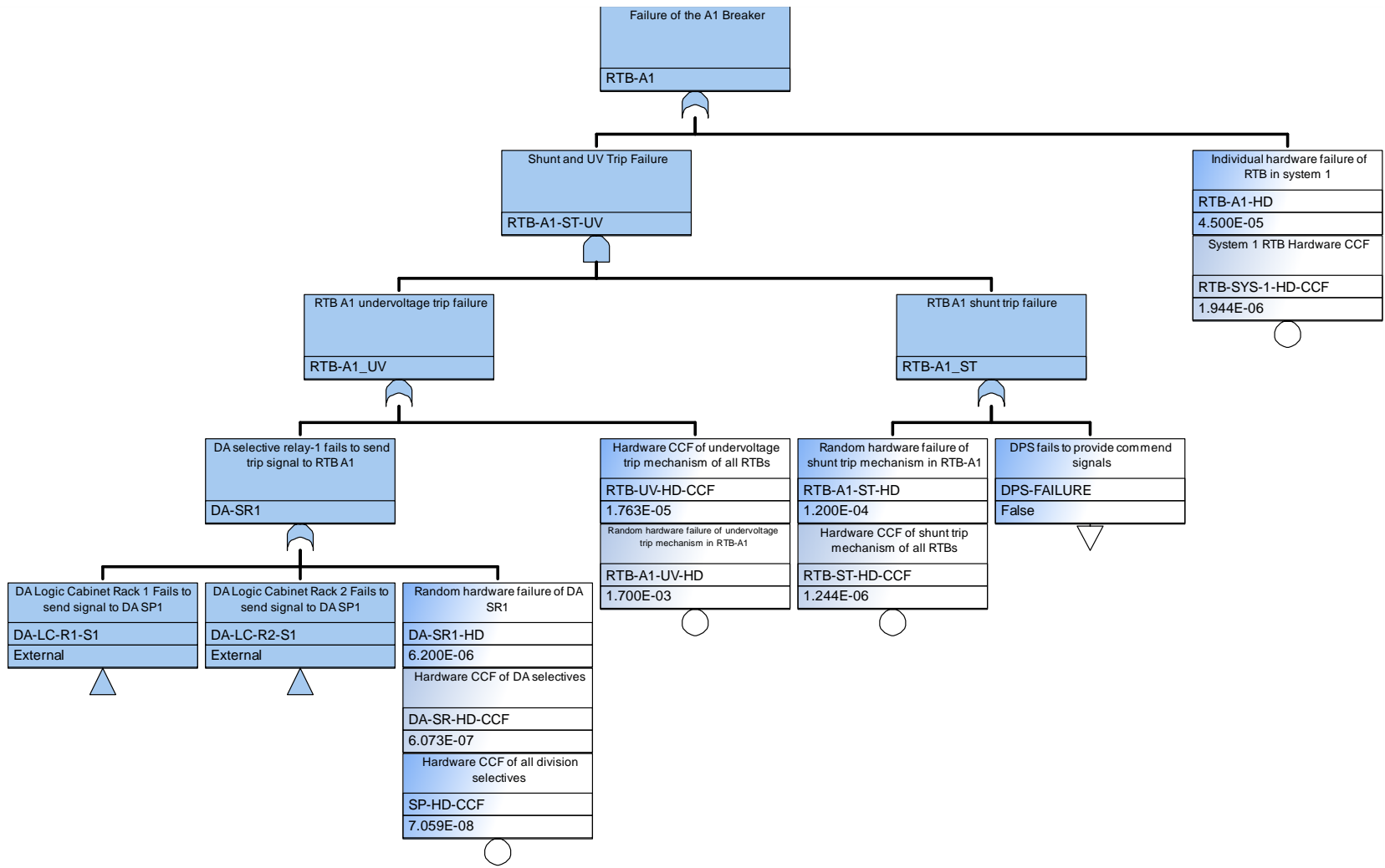

Figure 18. Transfer event of "Failure of A1 Breaker" of the integrated RTS-FT.

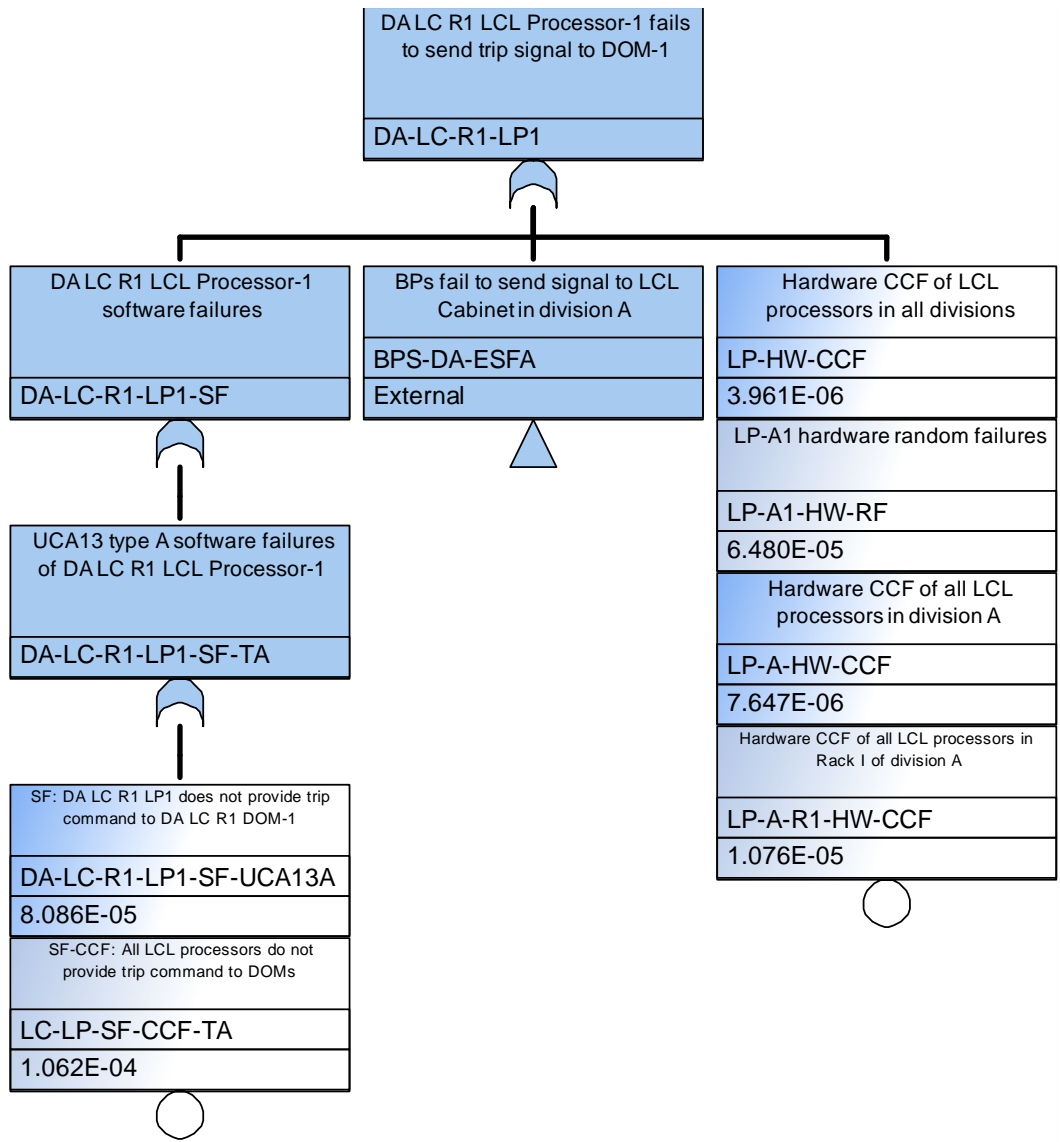

Figure 19. Transfer event of "DA LC R1 LCL Processor-1 fails to send trip signal to DOM-1" of the integrated RTS-FT.

Fault tree for "DA BP1 fails to send signal to DA LCL Cabinets" (DA-LC-BP1-DA), an OR gate with the following inputs:

- **DA LC BP1 software fails (affecting signal to DA LCLs)** — DA-LC-BP1-DA-SW
  - OR gate input: **UCA7-A software failures of DA LC BP1** — DA-LC-BP1-DA-UCA7-A
    - OR gate inputs:
      - DA BP1 does not provide command to DA LCL Cabinet when it's needed — DA-LC-BP1-DA-UCA7-A-SG — 5.591E-07
      - SW-CCF: DA BPs do not provide command to LCL Cabinet in all divisions when it's needed — DA-LC-BP-UCA-A-CCF — 1.062E-04
      - SW-CCF: All BPs do not provide command to LCL Cabinet in all divisions when it's needed — LC-BP-UCA-A-CCF — 8.030E-05
- **Hardware CCF of DA BPs** — DA-LC-BP-HW-CCF — 5.943E-06
- **Hardware CCF of all BPs** — LC-BP-HW-CCF — 2.187E-06
- **DA BP1 random hardware failure** — DA-LC-BP1-HW-RF — 4.000E-05
- **Failure of sensors to detect and/or provide initiating signal to DA** — DA-SEN-TRANS — False (transfer)

Figure 20. Transfer event of "DA BP1 fails to send signal to DA LCL Cabinets" of integrated RTS-FT.

**Appendix B:** PWR Event Trees for Consequence Analysis of DI&C Failures RTS (derived from [9]).

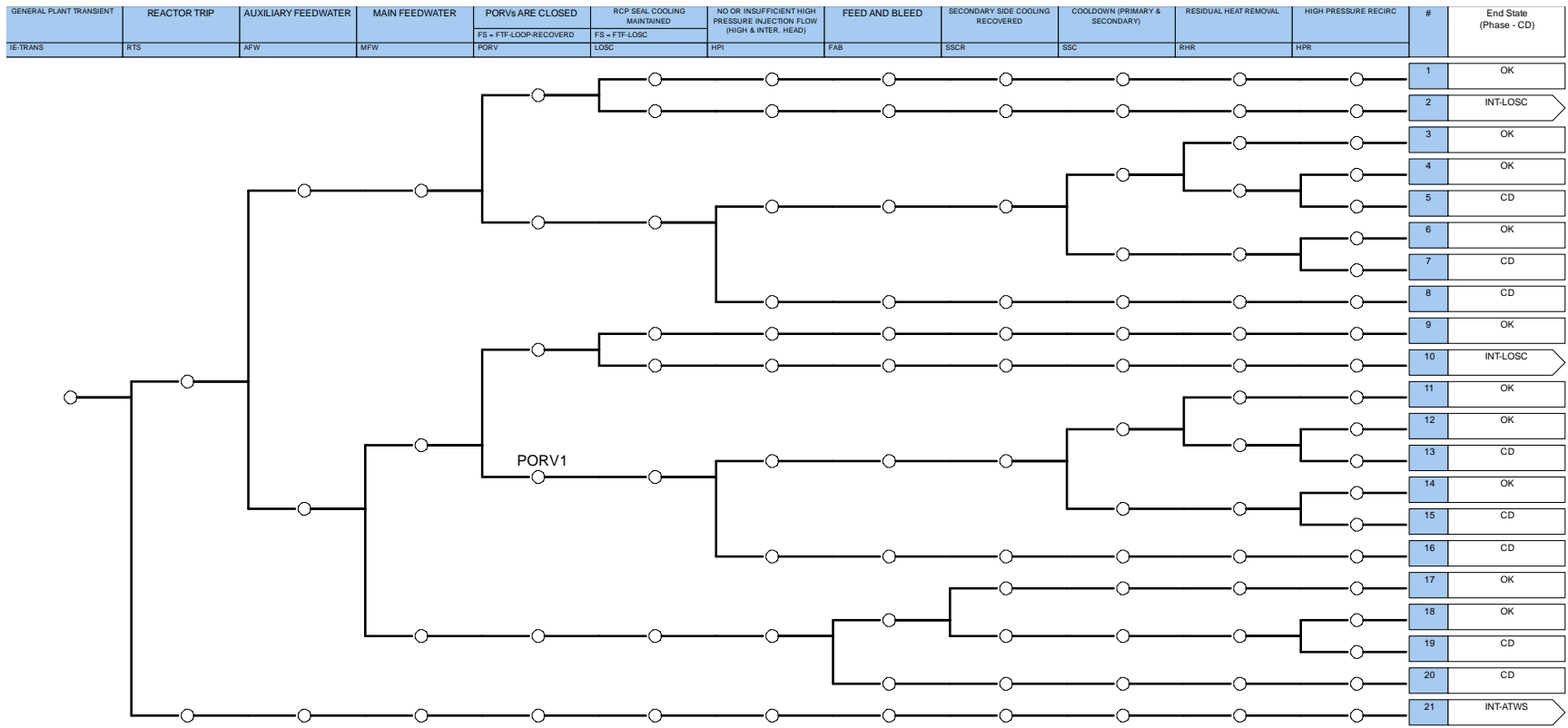

Figure 21. Generic PWR ET for general plant transient (INT-TRANS).

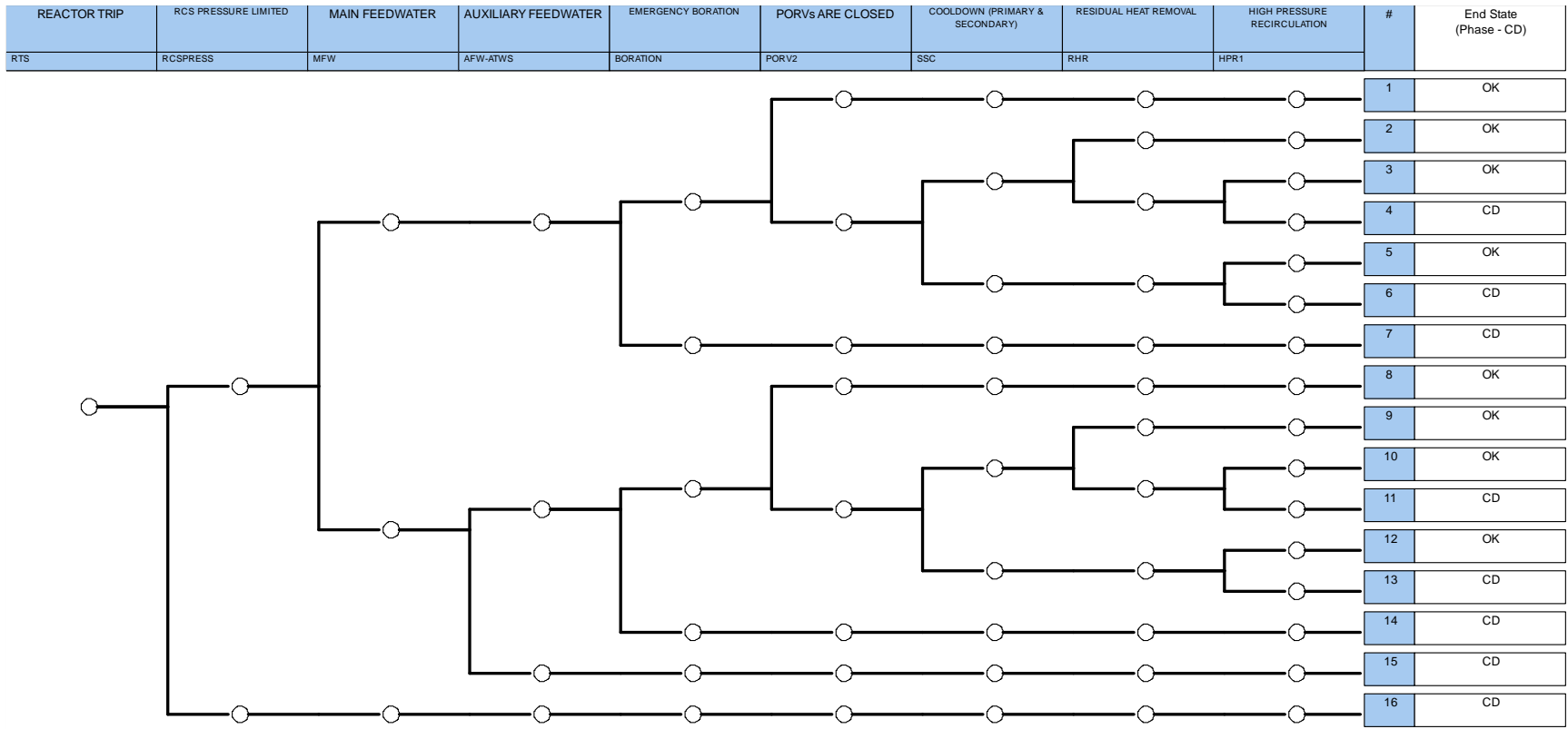

Figure 22. Generic PWR ET for INT-ATWS.

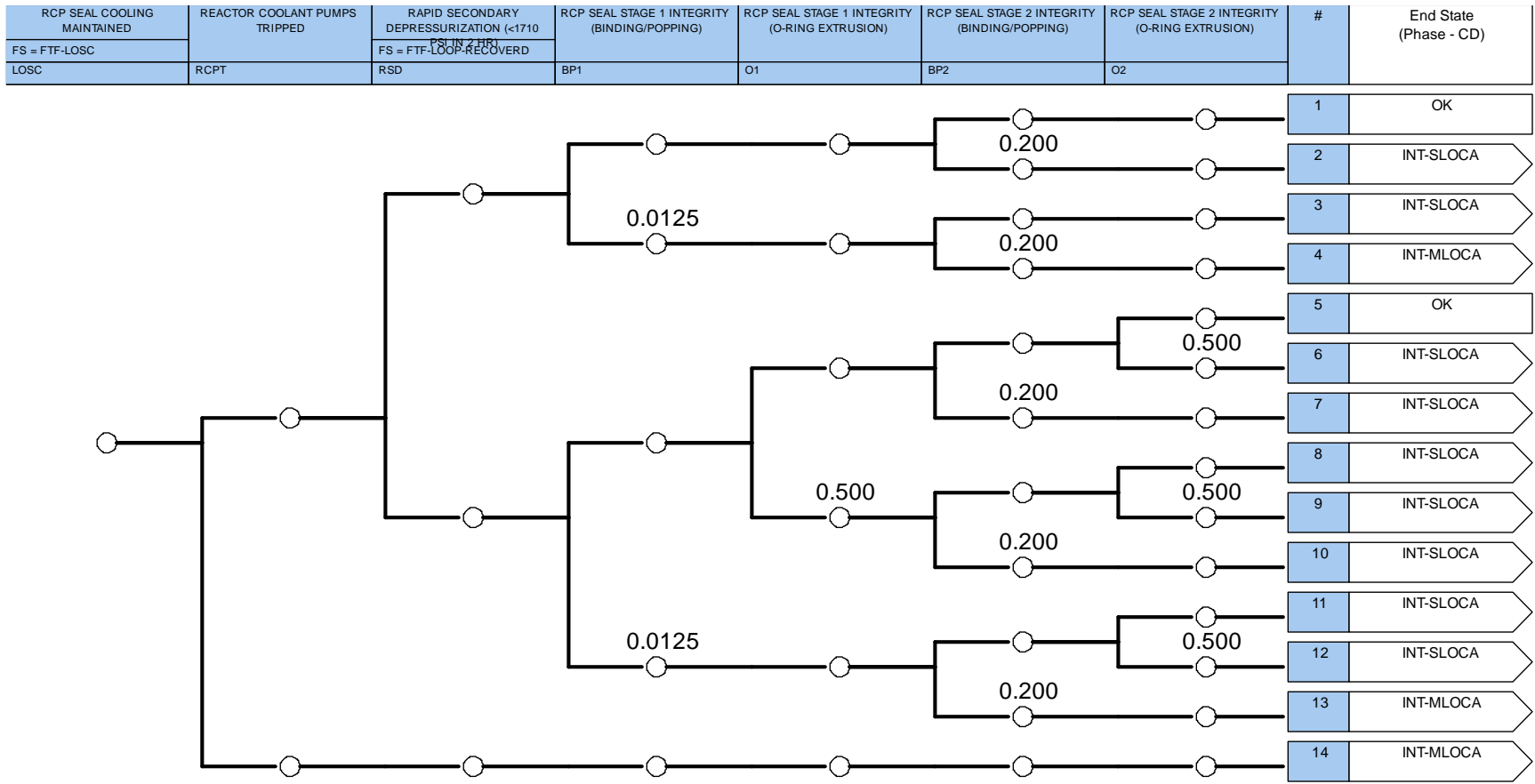

Figure 23. Generic PWR ET for loss-of-seal cooling (INT-LOSC).

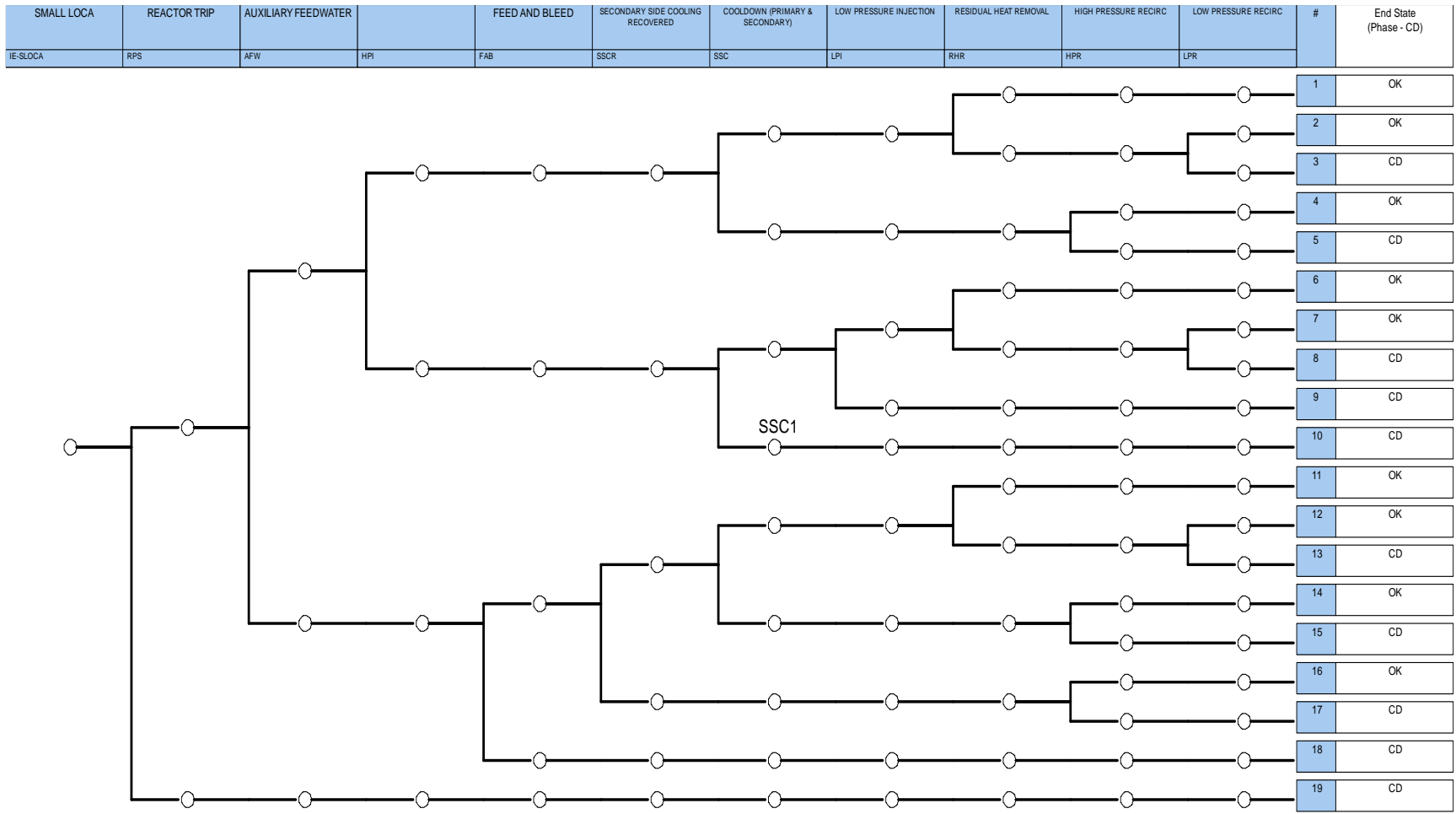

Figure 24. Generic PWR ET for small-break LOCA (INT-SLOCA).

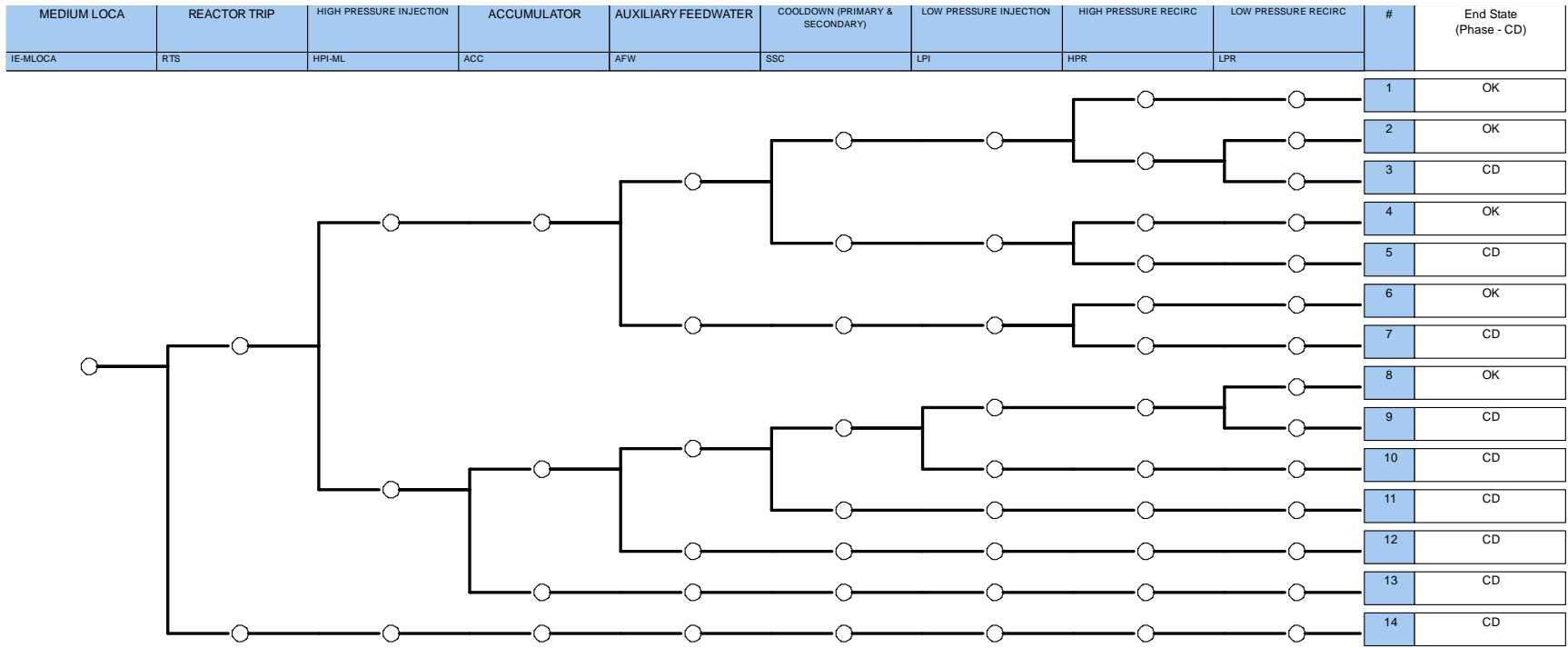

Figure 25. Generic PWR ET for medium-break LOCA (INT-MLOCA).